\documentclass[12pt]{iopart}
\usepackage[pdftex]{graphicx}
\usepackage{sidecap}
\usepackage{graphicx}
\usepackage{rotating}
\usepackage{iopams}
\usepackage{amssymb}
\usepackage{mathptmx}
\usepackage{amssymb}
\usepackage{epstopdf}
\usepackage{mathrsfs}
\usepackage{sidecap}
\usepackage{subfigure}
\usepackage{wrapfig}
\usepackage{comment}
\usepackage{float}
\usepackage{graphicx,pb-diagram,mathrsfs}
\usepackage{eso-pic}
\usepackage {graphicx}
\usepackage[numbers]{natbib}
\makeatletter

\bibpunct{}{}{,}{s}{}{,}

\begin{document}
\newcommand{\vv}{\mathbf}
\newcommand{\hdblarrow}{H\makebox[0.9ex][l]{$\downdownarrows$}-}

\title{A Vortex Filament Tracking Method 
for the Gross--Pitaevskii Model of a Superfluid}

\author{Alberto Villois$^1$, Giorgio Krstulovic$^2$, Davide Proment$^1$, Hayder Salman$^1$}

\address{1: School of Mathematics, University of East Anglia, Norwich Research Park, Norwich, NR4 7TJ, United Kingdom\\
2: Laboratoire J.L. Lagrange, UMR7293, Universit\'e de la C\^ote d'Azur, CNRS, Observatoire de la C{\^o}te d'Azur, B.P. 4229, 06304 Nice Cedex 4, France
}
\ead{A.Villois@uea.ac.uk}

\date{\today}


\begin{abstract}
We present an accurate and robust numerical method to track quantized vortex lines in a superfluid described by the Gross--Pitaevskii equation. By utilizing the pseudo-vorticity field of the associated complex scalar order parameter of the superfluid, we are able to track the topological defects of the superfluid and reconstruct the vortex lines which correspond to zeros of the field. 
Throughout, we assume our field is periodic to allow us to make extensive use of the Fourier representation of the field and its derivatives in order to retain spectral accuracy.
We present several case studies to test the precision of the method which include the evaluation of the curvature and torsion of a torus vortex knot, and the measurement of the Kelvin wave spectrum of a vortex line and a vortex ring. The method we present makes no a-priori assumptions on the geometry of the vortices and is therefore applicable to a wide range of systems such as a superfluid in a turbulent state that is characterised by many vortex rings coexisting with sound waves. This allows us to track the positions of the vortex filaments in a dense turbulent vortex tangle and extract statistical information about the distribution of the size of the vortex rings and the inter-vortex separations. In principle, the method can be extended to track similar topological defects arising in other physical systems.
\end{abstract}

\vspace{2pc}
\submitto{\JPA}

\section{Introduction}
Superfluids have been the subject of investigation since 1937 when it was first discovered that when
liquid Helium $ ^4$He is cooled below the lambda phase transition temperature of $T_\lambda=2.7K$, it undergoes a phase transition to a superfluid state \cite{donnelly1991}.
In 1995, a dilute, weakly interacting and ultracold atomic gas of bosons was cooled below a critical temperature giving rise to a new state of matter, the Bose--Einstein condensate (BEC). This BEC arises as a consequence of atoms undergoing a macroscopic occupation of the lowest energy state of the system.
Atoms in the lowest energy state have the unique property that they can behave collectively and undergo collective oscillations allowing them to display quantum mechanical effects on a macroscopic scale. Soon after the realization of the first BECs, the phenomena of superfluidity was also confirmed to arise in such systems \cite{pitaevskii2003bose}.
Among the many fascinating properties exhibited by superfluids, such as the ability of the fluid to flow without dissipation, and the phenomena of second sound in $^4$He, particular interest has been raised by the discrete and quantized nature of vortices in these systems. In superfluids, the circulation around each vortex is quantized in units of Planck's constant. These quantized vortices correspond to topological defects of the order parameter of the system where the vorticity is represented by a distribution of Dirac's $ \delta $-functions but where the density of the superfluid vanishes along the centreline of the vortex. In this regard, superfluids provide a naturally occurring system in which a vortex filament representation of the vorticity provides an excellent model rather than being an artificial mathematical idealization as in the case of a classical vortex. 


Given these observations, several models have been developed to describe the motion of vortices in superfluids. These include the vortex filament model based on the Biot-Savart law and one of its approximations commonly referred to as the local induction approximation\cite{Schwarz1985,Saffman1992}. In this work, we will focus on a microscopic description of a superfluid provided in the form of the Gross Pitaevskii (GP) equation which can be formally derived for a weakly interacting Bose gas in the limit of zero temperature. However, it has also served as a useful hydrodynamic model of a superfluid since it can qualitatively reproduce the dynamics of vortices in liquid superfluid $^4$He \cite{Roberts2001}.
The Gross--Pitaevskii model provides a natural framework in which to address many open questions regarding the dynamics of vortices in these systems. This includes physical processes that are involved during vortex reconnections \cite{koplik1993vrs, nazarenko2003, zuccher2012}, the nonlinear interaction and the decay of helical excitations of vortex lines in the form of Kelvin waves \cite{Krstulovic2012, 2013arXiv1308.0852P, diLeoniKW}, the interaction of sound with vortices \cite{Vinen2001}, and helicity considerations associated with the dynamics of superfluid vortices \cite{Scheeler2014, PhysRevE.92.061001, Brachet2016}. 
All of these aspects of the vortex dynamics are encountered within a turbulent vortex tangle \cite{nore1997, berloff2002ssn, yepez:084501}. It follows that in order to understand aspects of turbulence and the constituent elementary processes involved during the relaxation  of a superfluid vortex tangle, it is essential to be able to extract the location of the vortex filaments from high fidelity numerical simulations. An accurate and reliable numerical method is, therefore, needed for this purpose.

Tracking vortex lines of a complex wave-function in three spatial dimensions is in general a challenging problem. Among the range of different methods present in the superfluid literature we recall: standard interpolation techniques \cite{zuccher2012, Taylor2014}, low-density averages \cite{PhysRevA.69.053601, 2013arXiv1308.0852P}, two-dimensional Newton-Raphson method on planes corresponding to the Cartesian mesh \cite{Krstulovic2012}, and contour plots of the pseudo-vorticity field \cite{Rorai2014}. Many of these methods turn out to be either geometry dependent or not accurate enough to evaluate non-trivial geometrical quantities such as curvature,  torsion and small-scale Kelvin waves on any vortex configuration. 

The goal of the present work is to present a novel method for extracting vortex lines of the complex wave-function based on the Newton-Raphson method for finding the zeros of a given function \cite{NumRecipes} for a three-dimensional periodic domain. In order to demonstrate the method's accuracy and robustness, we will present a detailed validation using different test cases. Our aim is to demonstrate the potential broad applicability of the algorithm through consideration of different physical scenarios that are important in superfluids. Application of the method to address specific physical questions is deferred to future work.
The paper is organised as follows. In Section 1, we present the GP model, its hydrodynamic interpretation, and recall the key properties of quantized vortex solutions of this equation.
In Section 2, we describe how to implement the Newton-Raphson method to track vortices in two-dimensional complex wave-functions and the role played by the pseudo-vorticity field. 
In Section 3, we generalize the ideas presented for the 2D case to allow us to develop a
novel algorithm to track vortex lines of a three-dimensional wave-function by making use of the Newton-Raphson method in combination with the pseudo-vorticity field. In Section 4, we 
demonstrate our algorithm on a number of case studies consisting of different vortex configurations to test the accuracy and the robustness of the method. This includes a detailed evaluation of the curvature and torsion of a perfectly circular vortex ring and a torus knot, two quantities that are very useful to characterise the intrinsic properties of a vortex filaments. 
As a further test, we also evaluate the Kelvin wave spectra of a perturbed vortex line and vortex ring to illustrate how the algorithm can accurately extract information of a filament across a broad range of scales. Finally we demonstrate that the method is capable of tracking several vortex rings and even a dense vortex tangle. In Section 5, we draw our conclusions on the tracking method and explain how its implementation will complement our present knowledge of superfluid dynamics in the Gross--Pitaevskii model.  

\section{The Gross Pitaevskii model}
It is known that a three-dimensional system of identical bosons cooled below a critical temperature undergoes a phase transition, called Bose Einstein condensation, where the ground state becomes macroscopically occupied.
In the limit of zero temperature and assuming a dilute weakly interacting Bose gas, a classical partial differential equation modelling the dynamics of the system can be formally derived within the Hartree--Fock approximation. Specifically, the system can be described in terms of a complex wave-function $\psi$ representing the order parameter of the condensate within a mean-field approximation. The time evolution is governed by a nonlinear Schr\"odinger equation, more commonly known in this context as the Gross--Pitaevskii (GP) equation \cite{pitaevskii2003bose} given by
\begin{equation}\label{eq:GP}
i\hbar \partial_t\psi(\vv{x},t)=\left(-\frac{\hbar^2}{2m}\nabla^2 + g|\psi(\vv{x},t)|^2+ V(\vv{x}, t) \right)\psi(\vv{x},t).
\end{equation}
Here $g=4\pi \hbar^2 a /m $ is the coupling constant that arises from the assumed contact interaction potential and is expressed in terms of the s-wave scattering length, $a$, $ \hbar $ is the reduced Planck's constant, and $ m $ is the mass of the boson. 
$ V(\vv{x}, t) $ typically corresponds to an external confining potential which we will assume to be zero throughout this work. 
Equation (\ref{eq:GP}) conserves the total mass of the system defined as
 \begin{equation}
 M=m\int|\psi|^2d^3\vv{x},
\end{equation}
and the total energy
\begin{equation}\label{E}
 E=\int \left( \frac{\hbar^2}{2m}  |\nabla\psi|^2 + \frac{g}{2} |\psi|^4 \right) d^3\vv{x}.
 \end{equation}
In addition, for the case of an unbounded or a periodic domain, the total linear momentum
\begin{equation}\label{eq:Lin_mom}
 \vv{P}=\frac{\hbar}{2i}\int[\psi^*\nabla\psi-\psi\nabla\psi^*]d^3\vv{x},
\end{equation}
is also conserved. By introducing the Madelung transformation given by
\begin{equation}\label{eq:Madelung}
\psi(\vv{x},t)=\sqrt{\frac{\rho(\vv{x},t)}{m}}e^{i \frac{m}{\hbar} \phi(\vv{x},t)},
\end{equation}
the GP equation can be transformed into a coupled set of hydrodynamic like equations such that
\begin{eqnarray}\label{eqcont}
&&  \frac{\partial \rho}{\partial t}+\nabla\cdot(\rho\vv{v})=0, \\
\label{EulerQ}
&& \frac{\partial \vv{v}}{\partial t}+\vv{v}\cdot \nabla\vv{v}+\frac{g}{m^2}\nabla\rho-\frac{\hbar^2}{2m^2}\nabla\left[\frac{\nabla^2\sqrt{\rho}}{\sqrt{\rho}}\right]=0.
\end{eqnarray}
These equations describe the motion of a compressible, irrotational and barotropic fluid with a density corresponding to $\rho(\vv{x},t)=m |\psi(\vv{x},t)|^2$ and a velocity field equal to $\vv{v}(\vv{x},t)=\nabla \phi(\vv{x},t)$.

We remark that, even though the velocity is derived from a velocity potential, the equations can nevertheless admit vortex solutions. These solutions correspond to singular distributions of vorticity (i.e.\ point vortices in 2D or vortex filaments in 3D). To ensure the flow field maintains finite momentum and energy, vortex solutions are characterised by a depletion of the superfluid density in the vicinity of the vortex such that the density vanishes along the axis of the vortices. In order to support such vortex solutions, the velocity must be a non single-valued function of position. However, as seen from equation\ (\ref{eq:Madelung}), $\phi$ is related to the phase of the wavefunction. Therefore, in order to ensure that the wave function (\ref{eq:Madelung}) remains single-valued, the circulation can only take discrete values equal to 
\begin{equation}\label{eq:circ}
\Gamma =\oint_{\mathcal{C}}\vv{v}\cdot d\vv{l}= \oint_{\mathcal{C}} \nabla\phi \cdot d\vv{l} = s\frac{h}{m}.
\end{equation}
Here, $s\in \mathbb{Z}$ is the winding number and the circulation is, therefore, quantized in units of $h/m $. Such quantized vortices correspond to topological defects of the order parameter and have a characteristic core size set by the region where 
the density field drops to zero. This core size is set by the healing length $\xi=\hbar/\sqrt{2g\rho_{\infty}}$, where $\rho_{\infty}$ is the condensate density in the far-field away from the vortex core. The healing length provides an intrinsic length scale of the fluid that is set by balancing the kinetic energy (linear term) and the interaction energy (nonlinear term) in the GP equation. Given that vortices correspond to the set of zeros of the complex field $\psi$ (i.e.\ where the density vanishes), we require
\begin{equation}\label{eq:roots}
\boldsymbol{\Psi}(\vv{x})\equiv {\psi_r(\vv{x}) \choose \psi_i(\vv{x})}\equiv {\bf 0},
\end{equation}
with $\psi_r\equiv\Re e\psi$ and $\psi_i\equiv \Im m\psi$. Therefore, vortices  generically correspond to points in two dimensions and filaments in three dimensions. The problem of tracking quantized vortices thus reduces to finding the nodal lines of the wavefunction.


\section{Tracking vortices in two-dimensional complex fields}\label{2-d}
To numerically track quantized vortices in a two-dimensional complex field, we can make use of the root-finding routine based on the Newton-Raphson (NR) method. This approach has already been used to accurately track quantized vortices in GP simulations \cite{Krstulovic2012,Krstulovic2DVortices}.
In order to present our method, we will begin by recalling the key elements of the NR method for the vector function $\Psi: \mathbb{R}^2\rightarrow \mathbb{R}^2$ defined as in equation\ (\ref{eq:roots}). We will assume that we are able to find a good initial guess $ \vv{x}^g $ for the true position of the vortex, denoted by $ \vv{x}^v $, such that $|\Psi(\vv{x}^g)|^2<\epsilon \, |\Psi_{\infty}|^2 $, where $|\Psi_{\infty}|\equiv\sqrt{\rho_{\infty}/m} $ and $ \epsilon $ is assumed to be sufficiently small. We can then express $ \Psi(\vv{x}^v) \equiv 0 $ in terms of a Taylor-expansion of the function $ \Psi $ about the initial guess to obtain
\begin{equation}
\Psi(\vv{x}^v)=\Psi(\vv{x}^g) +J(\vv{x}^v) (\vv{x}^v-\vv{x}^g) + \mathcal{O}\left[(\vv{x}^v-\vv{x}^g)^2\right] = {\bf 0} \, ,
\end{equation}
where 
\begin{equation}\label{eq.Jac}
J(\vv{x})= \left( \begin{array}{cc}
  \partial_x \psi_r(\vv{x})  & \partial_y \psi_r(\vv{x}) \\
  \partial_x \psi_i(\vv{x})  & \partial_y \psi_i(\vv{x})
  \end{array} \right),
\end{equation}
is a Jacobian $ 2\times2 $ matrix. 
Assuming that the Jacobian is invertible when evaluated at $ \vv{x}^g $, equation\ (\ref{eq.Jac}) can be used to obtain a better approximation for the vortex position, $ \vv{x}^v $ given by
\begin{equation}\label{eq.New-Raph}
\vv{x}^v = \vv{x}^g -J^{-1}(\vv{x}^g)\Psi(\vv{x}^g) + \mathcal{O}\left[(\vv{x}^v-\vv{x}^g)^2\right].
\end{equation}
This procedure can in principle be iterated as many times as necessary (using the most recent evaluation of $\vv{x}^v$ as the new initial guess for each iteration) in order to converge to the exact location of the vortex. In practice, we will assume a reasonably converged solution when the condition $|\Psi(\vv{x}^g)| <\Delta \, |\Psi_{\infty}| $ is satisfied, where $\Delta $ is an arbitrarily small parameter which we typically set to be of the order of machine precision.
The advantage of this method is the fast quadratic convergence provided by the NR method but does have the drawback that convergence is not ensured if the initial guess, $ \vv{x}^g $, is far from the true solution, $ \vv{x}^v $. A further important requirement to ensure a reliable and accurate numerical method is the accurate evaluation of $ \vv{x}^g $ and the Jacobian $J(\vv{x}^g)$ at points that do not necessarily coincide with the grid where the field data is stored. We address these issues by setting a sufficiently small value of $\epsilon$ and by using a Fourier series expansion of our wavefunction to maintain full consistency with the spectrally accurate representation of our complex scalar field that is recovered from our pseudo-spectral numerical simulations.

We note that simply finding a root of $\psi$ is not enough to detect a quantized vortex since certain solitons or solitary wave excitations can also result in a zero density field. We, therefore, also check that the circulation is non-zero. 
For a classical fluid, the circulation can be evaluated from knowledge of the vorticity field $\boldsymbol{\omega}=\nabla \times \vv{v}$.
However, in the GP model, this quantity is identically zero everywhere except at the vortex points where it corresponds to a Dirac $\delta$-function. A related quantity in quantum fluids can be recovered from the density current
\begin{equation}\label{eq:density_curr}
 \vv{j}=\rho\vv{v}=\frac{\hbar}{2i} \left( \psi^*\nabla\psi-\psi\nabla\psi^* \right).
\end{equation} 
This allows us to define a pseudo-vorticity field as 
\begin{equation}\label{eq:curl_density_curr}
\boldsymbol{\omega}_{ps}=\frac{1}{2}\nabla\times \vv{j},
\end{equation}
and has the advantage that it remains regular even at the axis of the vortices.
Now writing $\psi_r=(\psi+\psi^*)/2 $ and $ \psi_i=(\psi-\psi^*)/(2i)$, it is possible to define equation\ (\ref{eq:curl_density_curr}) in terms of the gradients of the real and the imaginary parts of $\psi$:
\begin{eqnarray}\label{eq:pseudo_vort}
\boldsymbol{\omega}_{ps}
&=\frac{\hbar}{4i}\left[\nabla \psi^*\times \nabla\psi - \nabla \psi\times \nabla\psi^* \right] \nonumber \\
&=\hbar\nabla\left[ \frac{\psi+\psi^*}{2}\right] \times \nabla\left[ \frac{\psi-\psi^*}{2i}\right] \nonumber \\
&=\hbar\nabla \psi_r(\vv{x})\times \nabla \psi_i(\vv{x}).
\end{eqnarray}
The pseudo-vorticity encodes important information about a vortex. For example, in two dimensions, the wavefunction of a vortex of winding number $s=\pm1$ located at the origin can be simply expressed in polar coordinates $(r,\theta)$ as $\psi(r,\theta)=\sqrt{\rho(r)/m}\,e^{i s\theta}$. Substituting this into the expression for $\psi$ in equation\ (\ref{eq:pseudo_vort}), we obtain
\begin{equation}
\boldsymbol{\omega}_{ps}=s \frac{\hbar}{m} \frac{1}{2r}\frac{\partial \rho}{\partial r} \hat{\vv{k}},
\end{equation}
where $\hat{\vv{k}}$ is a unit vector perpendicular to the plane containing the point vortex. For a GP vortex, with winding number $s=\pm 1$, it is possible to show that the density vanishes quadratically at the vortex core but relaxes to a constant far from the vortex (at length scales exceeding the healing length $\xi$)\cite{pismen1999vortices}. It follows that the pseudo-vorticity is finite at the vortex core and vanishes outside. Furthermore, its sign determines the charge of the vortex.



To provide an example of the ability of the NR algorithm in tracking vortices in a complex field in two spacial dimensions, we simulated a turbulent superfluid state characterised by several vortices and density fluctuations (sound waves) obtained using a two-dimensional GP model. The computational domain considered was periodic and was discretised using a uniform grid with $256^2$ points and a resolution of $\Delta x=\Delta y=0.5\xi$. The GP equation was integrated in time using a standard pseudo-spectral algorithm with a split-step method. 
The initial condition was generated by imprinting a number of vortices onto the condensate. This was accomplished by taking a product of the single-vortex wave-function defined using a Pad{\'e} approximation as described in \cite{Berloff2004}. To impose the periodicity of the field, special care was taken in considering vortex images in the domain. Using this initial condition, we then allowed the vortices to evolve and interact for a sufficiently long time in order to obtain a realistic turbulent flow with vortices coexisting with sound waves.
 \begin{figure}[!htb]
\minipage{0.24\textwidth}
\subfigure[]{
  \includegraphics[width=\linewidth]{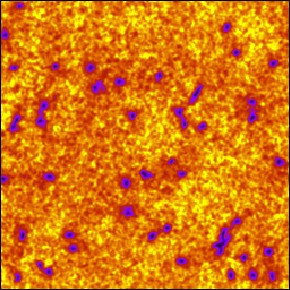} }
\endminipage\hfill
\minipage{0.24\textwidth}
\subfigure[]{
  \includegraphics[width=\linewidth]{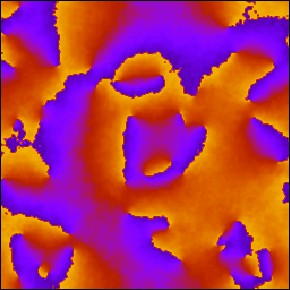}}
\endminipage\hfill
\minipage{0.24\textwidth}%
\subfigure[]{
  \includegraphics[width=\linewidth]{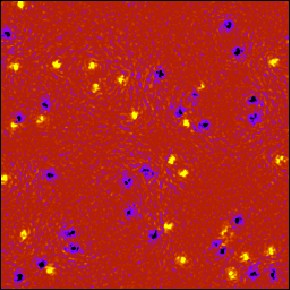}}
\endminipage
\hspace{0.cm}
\minipage{0.245\textwidth}%
 \subfigure[]{
  \includegraphics[width=\linewidth]{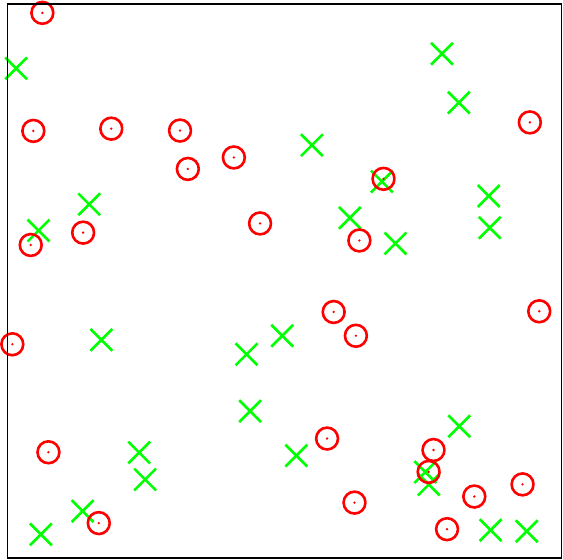}}
\endminipage
\caption{(Color online) Latest stage of the dynamics of 50 vortices in a two-dimensional GP model: (a) plot of the squared modulus of $\psi$ where dark colors represent depletions in $|\psi|^2 $ and light colors high values of $|\psi|^2$; (b) plot of phase field varying continuously from $-\pi h/m$, blue color, to $\pi h/m$, yellow color; (c) plot of the pseudo vorticity field, where yellow and blue points are respectively maxima and minima, while the filed vanishes in the red area; (d) tracked vortex positions, where red circles are vortices with negative circulations, while green crosses are vortices with positive circulations. 
\label{fig:2d_vort_ev}}
\end{figure}
In figures\ \ref{fig:2d_vort_ev}(a) and \ref{fig:2d_vort_ev}(b) we plot the squared modulus and the argument of $\psi$, respectively, at an intermediate time when a turbulent state has emerged. The vortices are clearly discernible in these plots and correspond to 
localised density depletions within a field that is otherwise dominated by density fluctuations extending over space. However, in regions where the vortices are tightly clustered together (see for instance the bottom right region), it is not immediately obvious what are the corresponding number of vortices in these regions.
In figure \ref{fig:2d_vort_ev}(c) we present the computed pseudo-vorticity field. As can be seen, this field clarifies not only the number of vortices present in tight clusters but also their charge. In figure \ref{fig:2d_vort_ev}(d) we plot the extracted vortex positions as computed using the NR technique, distinguishing between vortices characterized by positive (red circle) and negative (green cross) values of circulation.
We remark that the tracking parameters chosen here were $ \Delta = 10^{-13} $ and $ \epsilon=0.3 $. The average number of NR iterations were around 4-5 per vortex while the computational time involved was of the order of a few seconds on a standard desktop machine.

\section{Tracking vortex filaments in three-dimensional complex fields}
In contrast to the two-dimensional case where vortices appear in the form of point-like defects, in three spatial dimensions, quantized vortices correspond to filaments which can either form closed loops or end at domain boundaries. 
The configuration of the filaments can be arbitrarily complicated usually exhibiting non-trivial functional dependence on the local curvature and torsion\cite{Salman2013, KondaurovaVFMStat}. Aside from the complex vortex geometries that can arise, quantized vortices can also have a non-trivial topology by organizing themselves into knotted and/or linked structures\cite{Scheeler2014,maggioni:2010, PhysRevE.85.036306, 1742-6596-544-1-012022, Kleckner:2016yu}.

Given the much more diverse range of scenarios that can arise in 3D, tracking vortex filaments has remained a significant challenge in simulations of the Gross--Pitaevskii equation. In order to generalize our method to vortex filaments, we will firstly note that the
NR method cannot be directly applied to the function $ \Psi: \mathbb{R}^3\rightarrow \mathbb{R}^2$ since the Jacobian (\ref{eq.Jac}) would no longer be a square matrix. Alternatively, one might try to apply the method to $\psi$ defined on
planes orthogonal to the Cartesian coordinates. This approach would present difficulties in the case when the vortex filament is (almost) tangent to the plane since the determinant of equation\ (\ref{eq.Jac}) becomes almost singular thereby severely degrading the convergence and stability of the method. 
Another separate problem that arises in 3D is associated with how to recognise that different points in space correspond to zeros of the wavefunction belonging to the same vortex. This information concerning the connectivity of vortex filaments within a tracking algorithm presents a major issue when either a high density of vortex filaments are present in the computational domain or when two vortex filaments are extremely close to one another.
 
To circumvent these potential difficulties, we will make use of the pseudo-vorticity field defined by equation\ (\ref{eq:pseudo_vort}).
Considering that a vortex filament is a curve where both $\psi_r$ and $\psi_i$ have to vanish, it follows that projections of both $\nabla \psi_r$ and $\nabla \psi_i$ onto the curve must also vanish. 
These considerations reveal that the pseudo-vorticity field (\ref{eq:pseudo_vort}) is always tangent to the vortex filament.
Such a vector field evaluated in the vicinity of a filament not only allows us to identify a plane that is essentially orthogonal to the filament, but it also provides the connectivity information that allows us to determine which points belong to the same vortex.

The tracking algorithm we propose can now be summarised into the following steps (see also figure \ref{fig:Sketch} for an illustration):
\begin{figure}[t]
 \centering
      \includegraphics[scale=0.35]{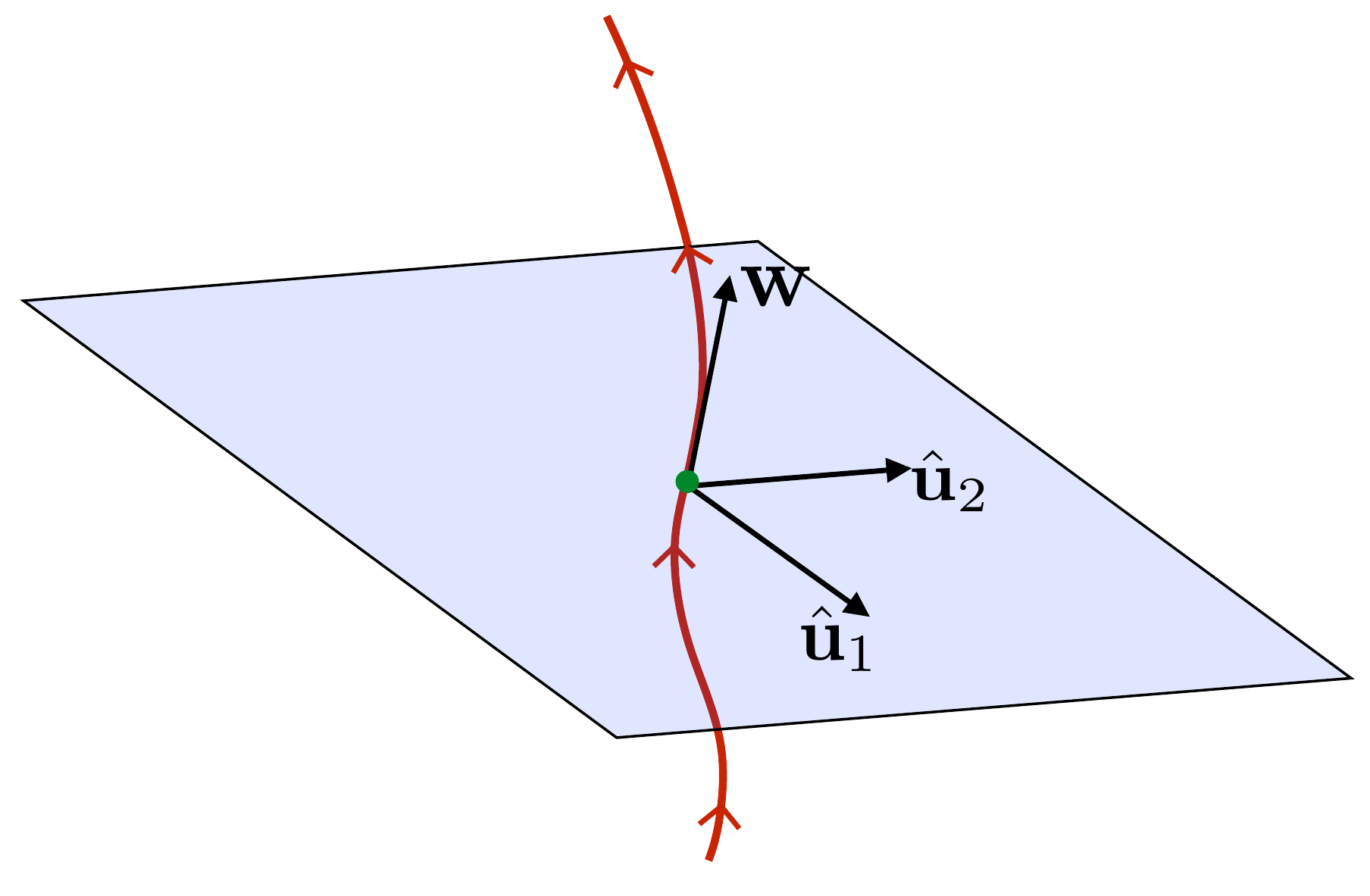}
 \caption{Sketch of the plane $\Pi$ where we develop the tracking.
\label{fig:Sketch}}
\end{figure}
\begin{itemize}
\item [1.] we begin with an arbitrary initial guess for a zero of $\psi$ which we denote by 
$\vv{x}^g $ such that $|\psi(\vv{x}^g)|^2 <\epsilon \, |\Psi_\infty|^2 $;
\item [2.] for $ \epsilon $ sufficiently small, $\vv{x}^g$ will be sufficiently close to the vortex filament. 
We can then evaluate the pseudo-vorticity field $\boldsymbol{\omega}_{ps}(\vv{x}^g)$ and assume that 
it points in a direction that is almost parallel to the vortex.
\item [3.] we define $\Pi$ as the plane that passes through the point $\vv{x}^g $ and orthogonal to the direction of $
\boldsymbol{\omega}_{ps}(\vv{x}^g) $. This allows us to identify an orthogonal basis $(\hat{\vv{u}}_1,\hat{\vv{u}}_2)$
for such a plane up to some arbitrary rotation.
\item [4.] we apply the NR method on the plane $\Pi$ obtaining
\begin{equation}
\vv{x}^v = \vv{x}^g-J^{-1}_\Pi (\vv{x}^g)\Psi(\vv{x}^g) + \mathcal{O}\left[(\vv{x}^v-\vv{x}^g)^2\right] \, ,
\end{equation}
where the $ 2\times2 $ Jacobian matrix projected on the plane $ \Pi $ is given by
\begin{equation}\label{Jac_gen}
J_\Pi(\vv{x})= \left( \begin{array}{cc}
 \nabla \psi_r(\vv{x})\cdot\hat{\vv{u}_1},   &   \nabla \psi_r(\vv{x})\cdot\hat{\vv{u}_2}\\
 \nabla \psi_i(\vv{x})\cdot\hat{\vv{u}_1}, &  \nabla \psi_i(\vv{x})\cdot\hat{\vv{u}_2}
  \end{array} \right).
\end{equation}
We note that by construction $ \vv{x}^v $ will always lie on the plane $ \Pi $.
As for the application of the NR method in the two-dimensional case, we now keep iterating over these steps until the condition $|\Psi(\vv{x}^v)| <\Delta  \, |\Psi_\infty| $ is satisfied;
\item [5.] we now interpret the point $ \vv{x}^v $ to belong to the $ j $'th vortex filament and store it as $ \vv{x}_i^{(j)}=\vv{x}^v $. The integer, $ i $, is set to one at the beginning of the tracking for each vortex filament and is incremented for each consecutive point extracted along a given filament. In order to identify the next point, $ \vv{x}^{(j)}_{i+1} $, along the filament, we make use of the fact that the pseudo-vorticity field is tangent to the curve. 
We, therefore, set our next guess to correspond to
\begin{equation}\label{eq:rule}
\vv{x}^g=\vv{x}^{(j)}_{i}+\zeta \hat{\boldsymbol{\omega}}_{ps}(\vv{x}_{i}),
\end{equation}
where $\zeta$ is an arbitrary small parameter and $ \hat{\boldsymbol{\omega}}_{ps}(\vv{x}_{i}) $ is the pseudo-vorticity vector normalised to unity;
\item [6.] we return to step ($ 2. $) in our algorithm unless $ i>1 $ and one of the following closing conditions for the $ j $'th vortex filament is satisfied:  
\begin{itemize}
\item[a)] the Euclidean distance $ d_{1k}^{(j)} = |\vv{x}_1^{(j)}-\vv{x}_k^{(j)}| $ is much less than the parameter $ \zeta $ (we typically set $ d_{1k}^{(j)} < \zeta/3 $).
Indeed, for values of $\zeta$ that are small relative to the local radius of curvature of the filament, we can assume that the arclength between two consecutive points is approximately equal to $\zeta$. 
Hence, when the distance $d_{1n}^{(j)} $ becomes much smaller than $ \zeta $, we an assume that the $ j $'th vortex filament to be closed and identify it with a {\it vortex loop};
\item[b)] the point $ \vv{x}^{(j)}_i = (x_i^{(j)}, y_i^{(j)}, z_i^{(j)}) $ on the $ j $'th vortex filament is identified with a {\it vortex line} if it satisfies the conditions $ \vv{x}^{(j)}_i \simeq (x_1^{(j)}\pm L_x,y_1^{(j)},z_1^{(j)})$ or $ \vv{x}^{(j)}_i \simeq (x_1^{(j)},y_1^{(j)}\pm L_y,z_1^{(j)})$ or $  \vv{x}^{(j)}_i \simeq (x_1^{(j)},y_1^{(j)},z_1^{(j)}\pm L_z)$ where $L_x$, $L_y$ and $L_z$ denote the dimensions of the periodic domain along the three coordinate directions, respectively. We note that due to the assumed periodicity of the field, points such as $(x_1^{(j)}\pm L_x,y_1^{(j)},z_1^{(j)})$ do not need to lie within the computational domain;
\end{itemize}
\item [7.] the $ j $'th vortex loop or line is then stored and we return to step ($1.$) to search for the next vortex filament.
In order to avoid re-tracking the same filament multiple times, we make use of a Boolean matrix, initially having all values set to zero, with same size of the grid on which $\psi$ is discretized.
After tracking the $ j $'th vortex filament, we set the Boolean matrix equal to unity on all the grid points close to it.
The initial guess for the $ (j+1) $'th vortex filament is then explored within the remaining volume grid points that correspond to zero entries of the Boolean matrix. 
\end{itemize}



\section{Case studies}
In order to demonstrate the accuracy and robustness of the tracking method described above, we will present several test cases corresponding to different vortex configurations.
We will initially focus on systems containing a single vortex filament in the form of a vortex loop. We will begin with the simplest possible geometry, a perfectly circular vortex ring before proceeding to track a topologically non-trivial vortex such as a torus vortex knot. To test the accuracy of the method, we compute geometrical quantities such as the coordinates, curvature and torsion of the filaments and compare these against their respective theoretical values.
Both cases are evaluated in a computational domain consisting of a grid of $ 128^3 $ points and a resolution $\Delta x=\Delta y=\Delta z=0.5\xi$.

We begin by considering a ring with a radius $R=4\xi$ moving along the $z$-coordinate direction. 
To set the initial condition, we use the analytical expressions obtained using a Pad\'{e} approximant by Berloff\cite{Berloff2004} for a two-dimensional vortex on the $x$-$y$ plane centred at $(0,R)$ and then rotating such a plane around the $z$-axis. 
This produces a vortex ring that is located along the curve given by
\begin{equation}
\vv{s}(\sigma)=(R\cos\sigma,R\sin\sigma, 0) \, ,\label{Eq:Ring}
\end{equation}
where the parameter $\sigma\in [0,2\pi)$.
After tracking the filament, we computed the geometrical distance between the tracked vortex points and the exact vortex filament to evaluate the error. This gave a maximum value of the order of $\sim 10^{-7}$ thus verifying the precision to which we are able to track the vortices.
To further check the accuracy of the method, we calculated the curvature $\kappa$ and torsion $\tau $, that require the evaluation of high order derivatives of the positions of the filament with respect to the parametrisation of the curve. These can be obtained using the expressions
\begin{equation}
\kappa=\frac{|\vv{s}'\times \vv{s}''|}{|\vv{s}'|^3},\qquad  \tau ={{\left( \vv{s}' \times \vv{s}''\right)\cdot \vv{s}'''} \over |\vv{s}'\times \vv{s}''|^2},
\end{equation} 
where $\vv{s}'$,$\vv{s}''$ and $\vv{s}''$ represent the derivatives with respect to the parameter $\sigma$. 
From the tracked points, we estimated the constant curvature of the ring to be $\kappa=(0.25\pm 0.2 \times 10^{-5}) \xi $ and the torsion to be $\tau=(0\pm 10^{-4}) \xi $.
These values are in excellent agreement with the theoretical values for the curvature and torsion given by $\kappa_{th}=1/4 \xi $ (inverse of the ring radius) and $\tau_{th}=0$, respectively. We remark that accurate evaluation of these quantities is necessary in many problems involving vortex dynamics. For example, curvature and torsion provide an intrinsic description of the geometry of a vortex filament that could be used to identify soliton like excitations\cite{Hasimoto1972}. On the other hand, accurate evaluation of the torsion is relevant to the study of helicity for vortex filaments\cite{Ricca1992,Hietala2016}. Therefore, the ability to extract these quantities will be invaluable in understanding fundamental aspects of the dynamics of superfluid vortices.

\begin{figure}
 \begin{minipage}[b]{0.48\textwidth}
    \vspace{0pt}
    \centering
    \subfigure[]{
      \includegraphics[scale=0.40]{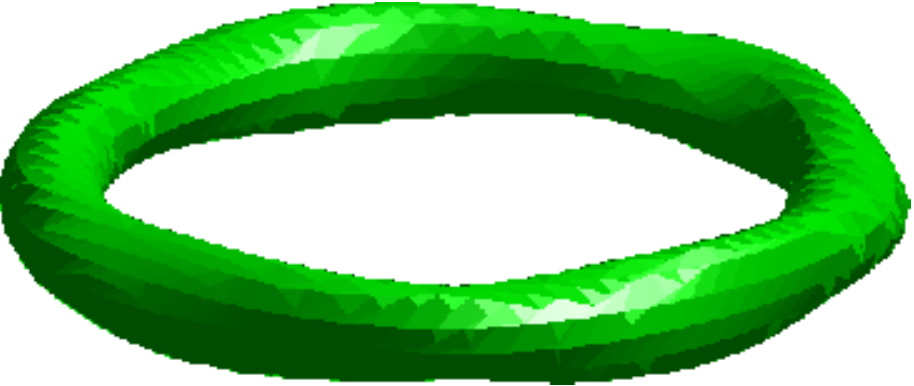}}
  \end{minipage}
 \begin{minipage}[b]{0.48\textwidth}
    \vspace{0pt} 
    \centering
    \subfigure[]{
      \includegraphics[scale=0.40]{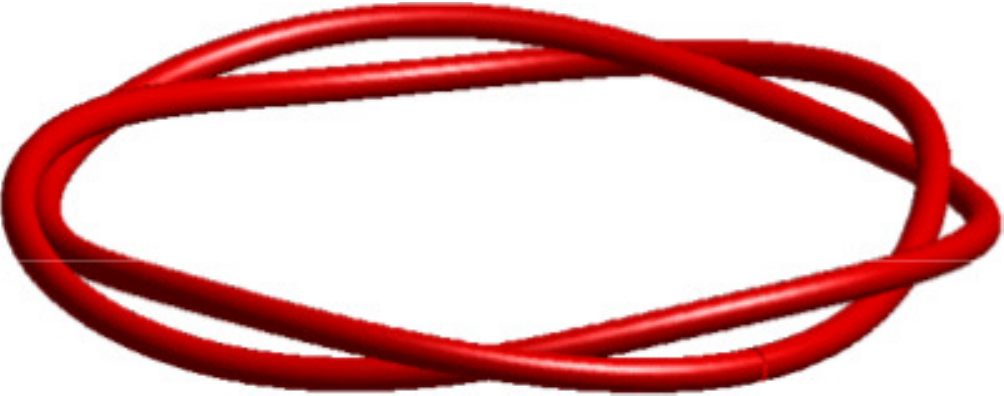}}
  \end{minipage}
 \begin{minipage}[b]{0.48\textwidth}
    \vspace{0pt}
    \centering
    \subfigure[]{
      \includegraphics[scale=0.28]{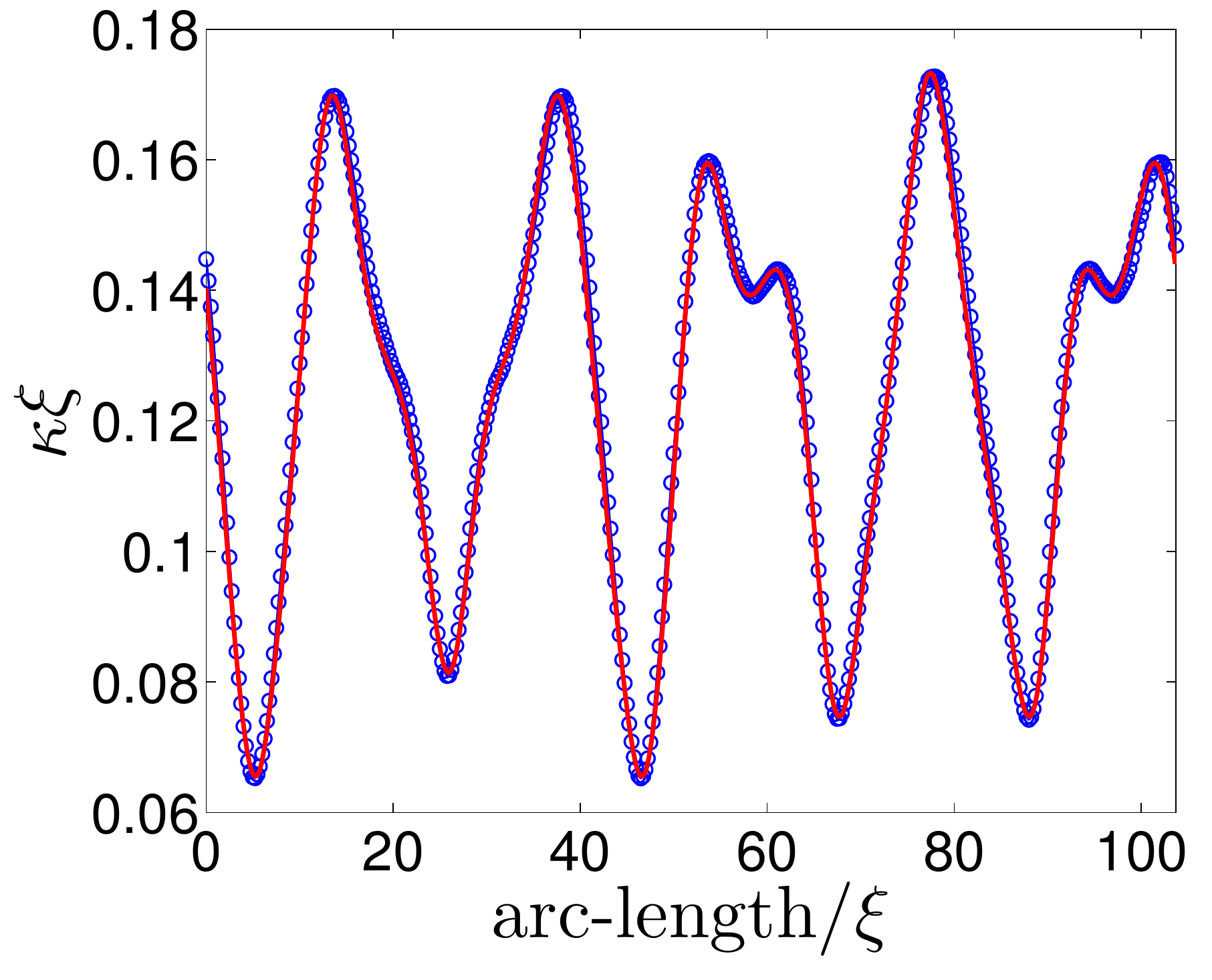}}
  \end{minipage}
  \hspace{0.1cm}
 \begin{minipage}[b]{0.48\textwidth}
    \vspace{0pt} 
    \centering
    \subfigure[]{
      \includegraphics[scale=0.28]{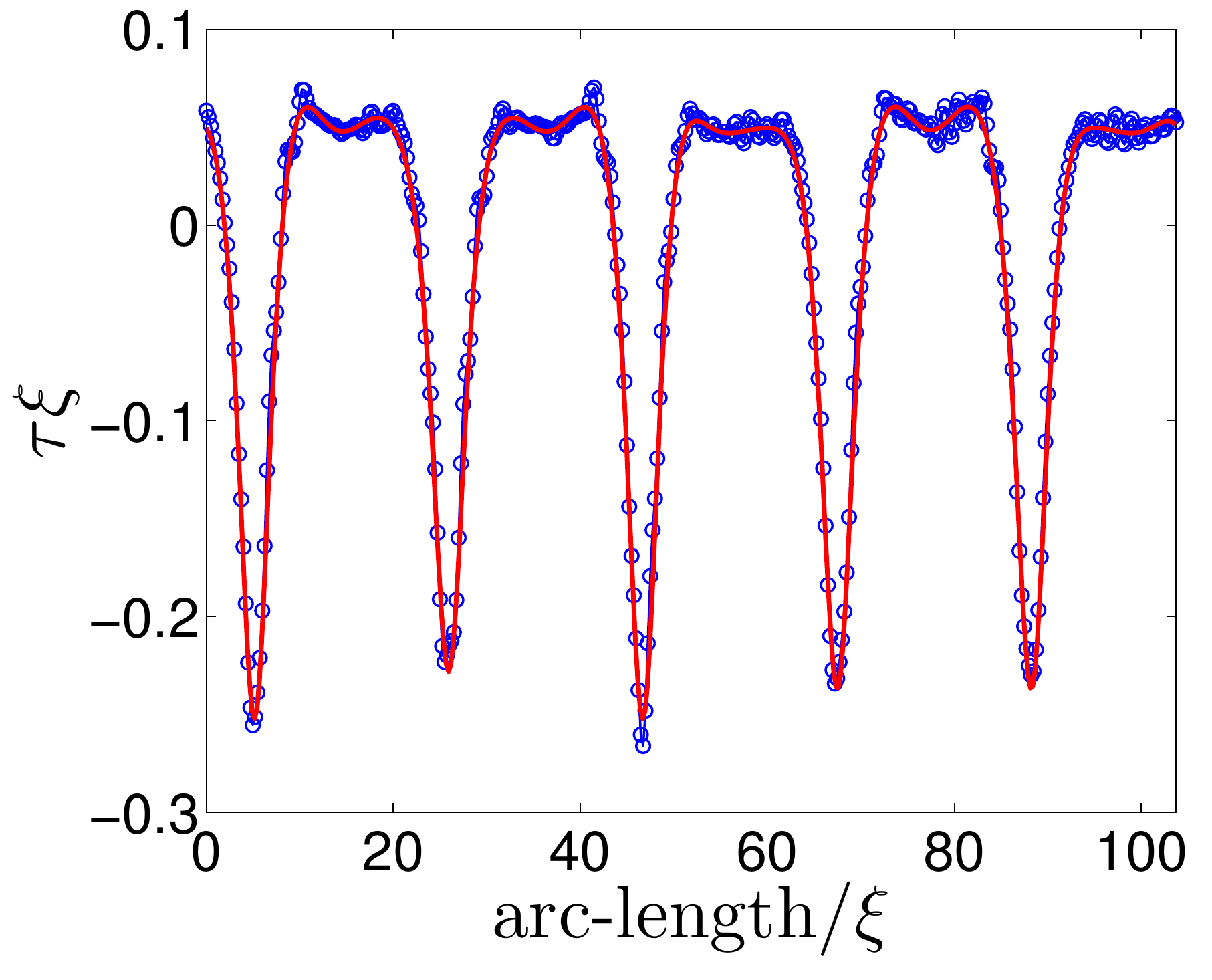}}
  \end{minipage}
 \caption{(Color online) (a) Iso-surface plot corresponding to $|\psi|^2=0.1 \, |\Psi_\infty|^2 $ for a knot as described in Appendix \ref{App:Knot}; (b) Plot of the tracked vortex filament where the vortex line has been rendered as a vortex tube to clearly demonstrate the non-trivial topology; (c) Plot of the curvature and (d) Plot of torsion versus arc-length. The numerical data (blue circled line) has been superimposed on the theoretical predictions (red line).
\label{fig:Knot}}
\end{figure}

Having demonstrated the algorithm on a simple vortex ring, we will now consider a vortex with a non-trivial topology: a torus vortex knot. The study of vortex knots in superfluids has attracted much interest in recent years as it provides an ideal paradigm to address questions related to helicity conservation in superfluids \citep{Brachet2016,PhysRevE.85.036306,Scheeler2014}. 
A torus knot $\mathbb{T}^{p,q}$ is a closed curve built on the surface of a torus having toroidal and poloidal radii $ R_0 $ and $ R_1 $ respectively that has been twisted $p$-times around the toroidal axis and $q$-times around the poloidal axis, where $p$ and $q$ are co-prime integers.
Its parametrisation ${\bf s(\sigma)}=(s_1(\sigma),s_2(\sigma),s_3(\sigma))$ is given by 
\begin{eqnarray}
&s_1(\sigma)=\left[ R_0+R_1\cos(p\sigma) \right] \cos(q\sigma),\\
&s_2(\sigma)=\left[ R_0+R_1\cos(p\sigma) \right]\sin(q\sigma),\\
&s_3(\sigma)=R_1\sin(p\sigma),\label{Eq:Knot}
\end{eqnarray}
where $\sigma\in[0,2\pi)$.
The wave-function containing a vortex knot $\mathbb{T}^{p,q}$ is obtained following the approach developed in \cite{PhysRevE.85.036306}. In order to enforce the periodicity of the field, we perform a coordinate transformation that is an approximation of the identity in the periodic domain. In \ref{App:Knot} we provide the explicit formulae used for the knot in the periodic domain.
As an example, we consider the torus knot $\mathbb{T}^{2, 5}$ build on a torus with a toroidal radius $R_0=8\xi$ and a poloidal radius $R_1=\xi/2 $.
In figure \ref{fig:Knot}(a) we plot the iso-surfaces of the density field of the wave-function corresponding to $ |\psi|^2=0.1  \, |\Psi_\infty|^2 $. 
For a knot with such a small poloidal radius (smaller than $\xi$), an iso-surfaces of the density does not allow us to clearly distinguish its geometry from a vortex ring. However, after tracking the filament, its topology becomes apparent, as shown in figure\ \ref{fig:Knot}(b).
In figures\ \ref{fig:Knot}(c) and \ref{fig:Knot}(d) we present a comparison between the numerically computed and analytical expressions for the curvature and torsion, respectively. The numerical results (shown as a blue circled line) coincide with the theoretical predictions 
with only a small amount of noise observed in the torsion that is caused by the way the initial condition for the wavefunction is generated.

\begin{figure}
 \begin{minipage}[b]{0.48\textwidth}
    \vspace{0pt}
    \centering
    \subfigure[]{
      \includegraphics[scale=0.30]{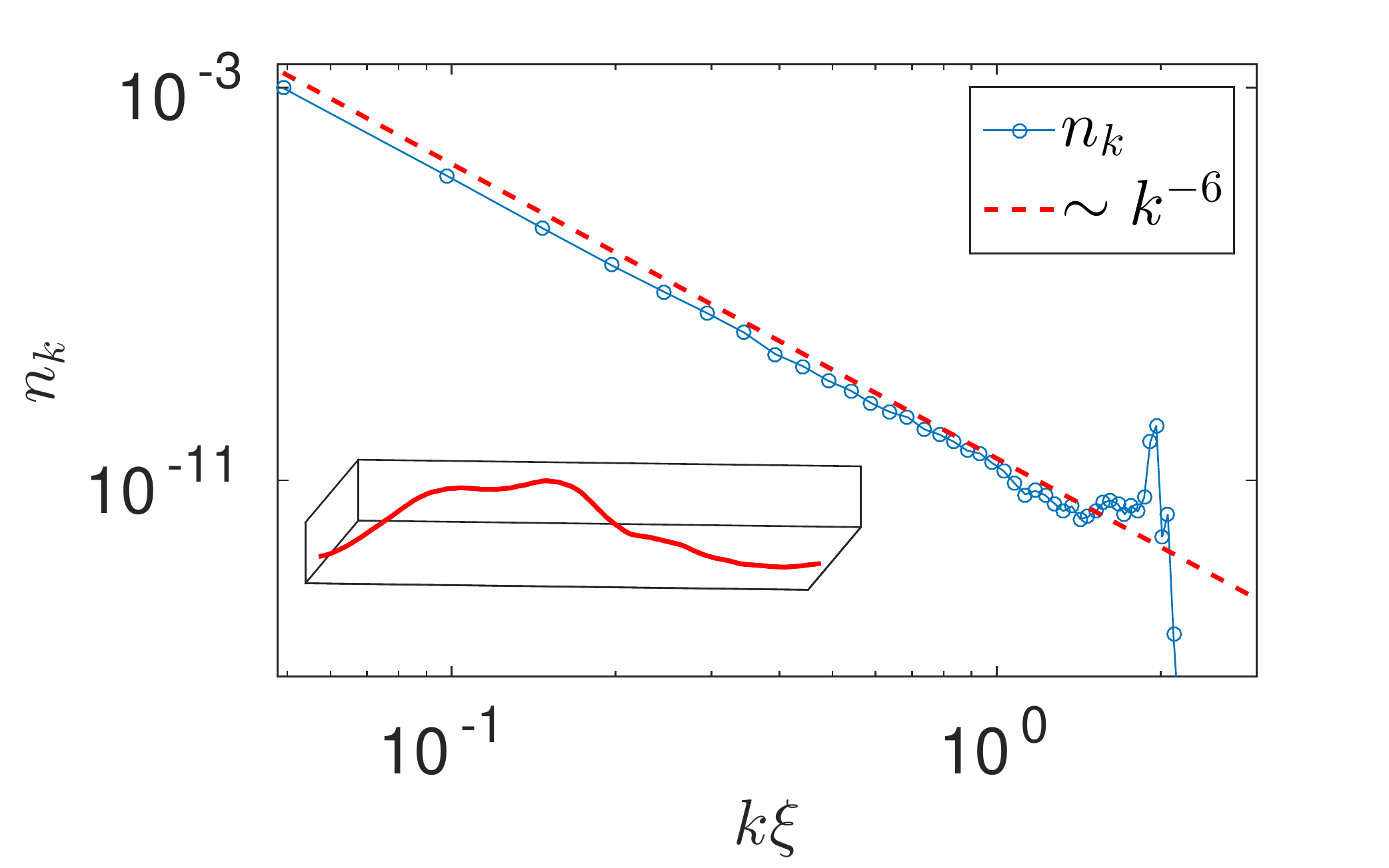}}
  \end{minipage}
  \hspace*{0.1cm}
 \begin{minipage}[b]{0.48\textwidth}
    \vspace{0pt} 
    \centering
    \subfigure[]{
          \includegraphics[scale=0.30]{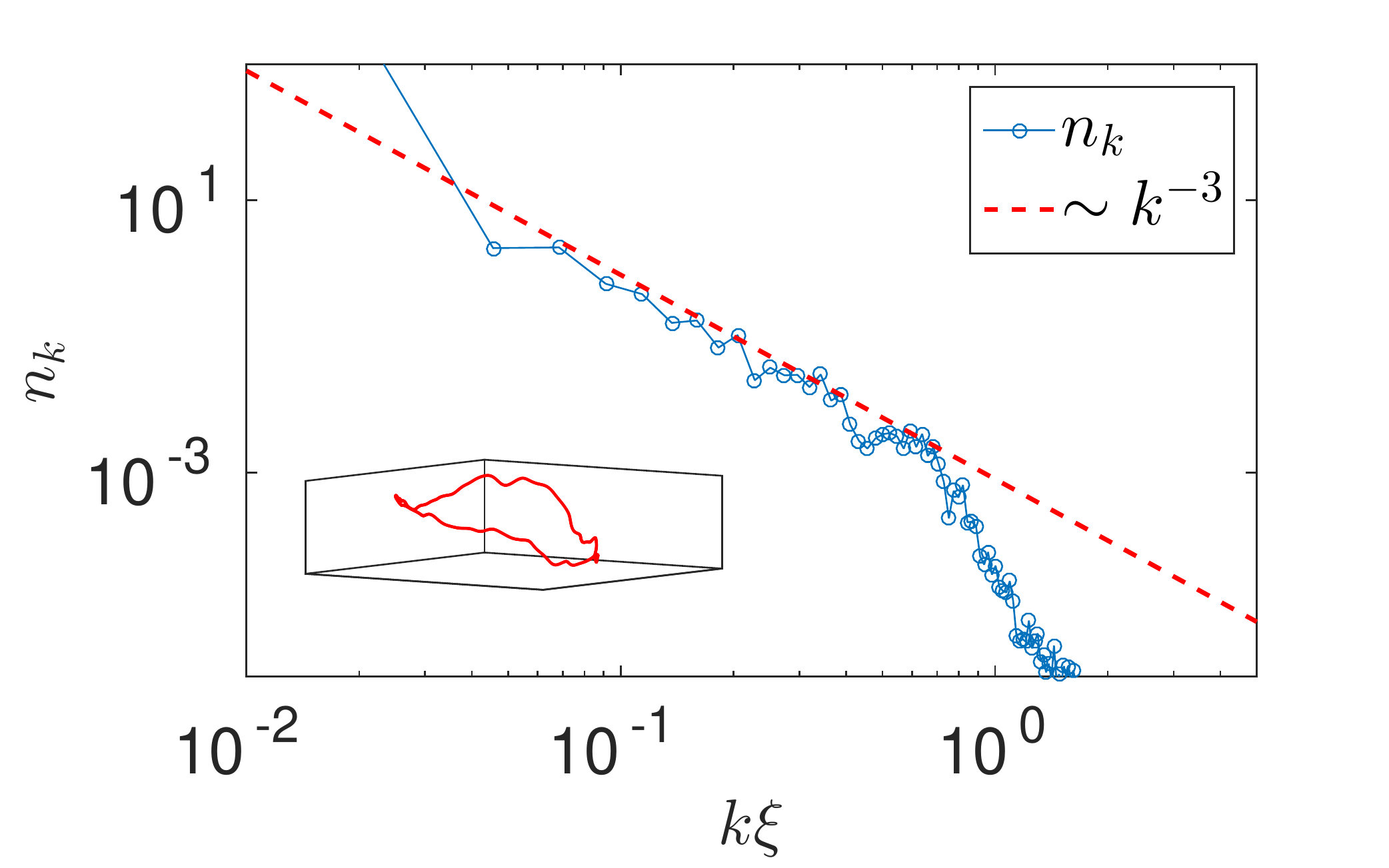}}
  \end{minipage}
  \caption{(Color online) (a) Log-log plot of the KW spectrum induced on a vortex line and (b) on a vortex ring, where a power law (red line) has been introduced for comparison with the numerical data (blue circled line). The insets show both the perturbed vortex line and the vortex ring.
\label{fig:spectra}}
\end{figure}

In order to illustrate the applicability of the algorithm to a problem containing multiple length scales, we now demonstrate
how it can be reliably used to detect small oscillations on a vortex filament. 
These excitations, called Kelvin waves (KWs), correspond to helical waves propagating along vortex filaments.
KWs and their energy spectra have been the object of investigation for some time since their nonlinear interactions is believed to provide a key mechanism in quantum turbulence to transfer energy down to length scales of the order of the vortex core where it can be dissipated through phonon emission. An accurate measurement of the KWs and their spectra on an almost straight vortex line using the two-dimensional NR method has already been performed by Krstulovic \cite{Krstulovic2012}. 
However, the tracking technique described in that work relied on a-priori knowledge of the vortex line configuration, namely that the vortex line was almost orthogonal to planes where the NR method is applied. 
Since our method is completely independent of the filament's configuration as well as orientation relative to the computational grid, we are able to measure not only the KW spectrum of a vortex line but also the KW spectrum of a vortex ring.
To create a wave-function characterised by KWs on a straight vortex line that is aligned along the $ z $-axis, we first consider the wave-function of the straight vortex line and then shift the $ x-y$ planes accordingly (for more details refer to \cite{Krstulovic2012}). KWs on a ring lying on the $x-y$ plane are imposed in a similar way by using the Pad\'e approximation with a perturbed radius together with a small $z$-component.
To test the ability of the algorithm to capture small amplitude KWs, we impose steep KW spectra corresponding to $ n(k) \sim k^{-6} $ and $ n(k) \sim k^{-3} $ on the vortex line and the ring, respectively. To smooth out any possible strong gradients in the field arising from the shifts used to construct the initial condition, we integrate our initial conditions over a short duration within the GP equation.
The insets of figures \ref{fig:spectra}(a) and \ref{fig:spectra}(b) illustrate the vortex line and the ring, respectively, with their corresponding KW excitations. The KW spectra are evaluated using the Fourier transform of the waves propagating along the straight line or ring. Details for obtaining the spectra are described in \ref{App:KW}.
In figures\ \ref{fig:spectra}(a) and \ref{fig:spectra}(b) we present the numerically computed KW spectra (blue circled line) together with the imposed scalings on the KWs that are shown as a red dashed line. The results clearly show that the spectra are reproduced by the algorithm down to the vortex core size. Note that whereas for the KWs on a straight line the spectrum is uniquely defined, on a ring KWs are defined as the perturbation of a smooth ring. To obtain the smooth ring, a convolution with some kernel is needed. The large scales of its spectra are thus kernel dependent, but small scales are independent (see \ref{App:KW} for further details).

\begin{figure}
 \begin{minipage}[b]{0.48\textwidth}
    \vspace{0pt}
    \centering
    \subfigure[]{
      \includegraphics[scale=0.15]{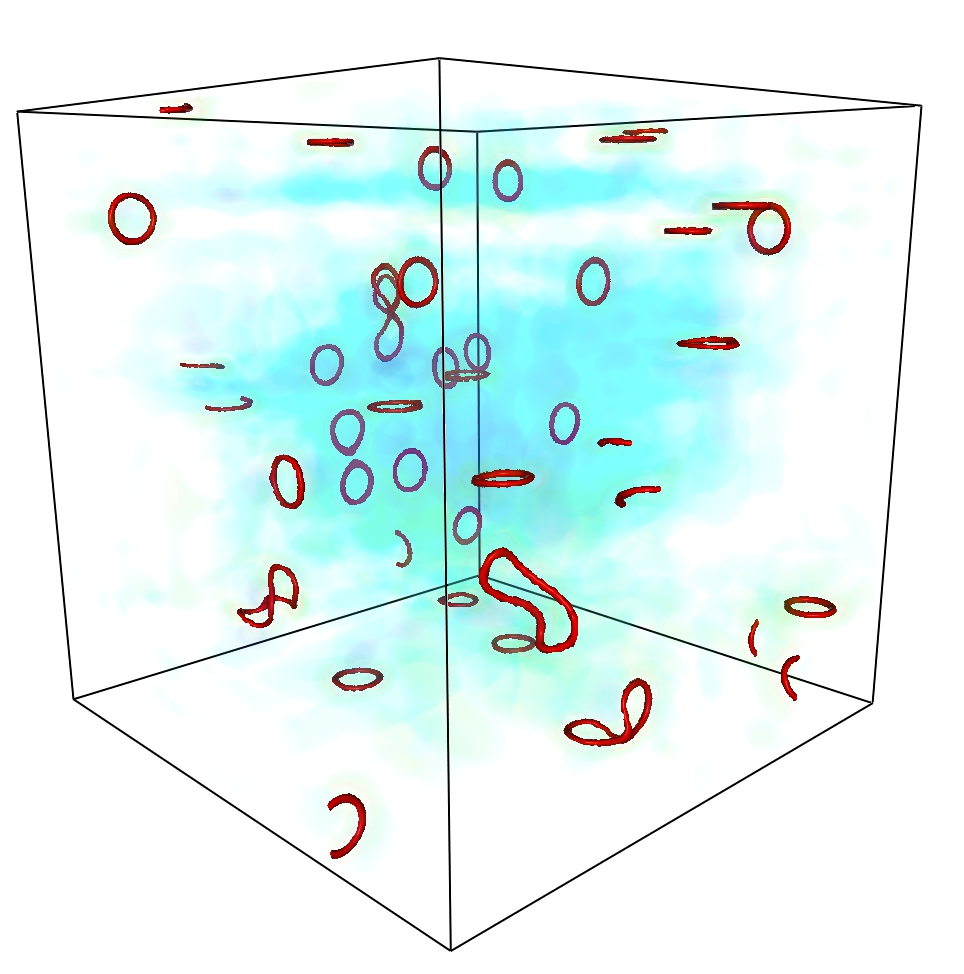}}
  \end{minipage}
 \begin{minipage}[b]{0.48\textwidth}
    \vspace{0pt} 
    \centering
    \subfigure[]{
          \includegraphics[scale=0.22]{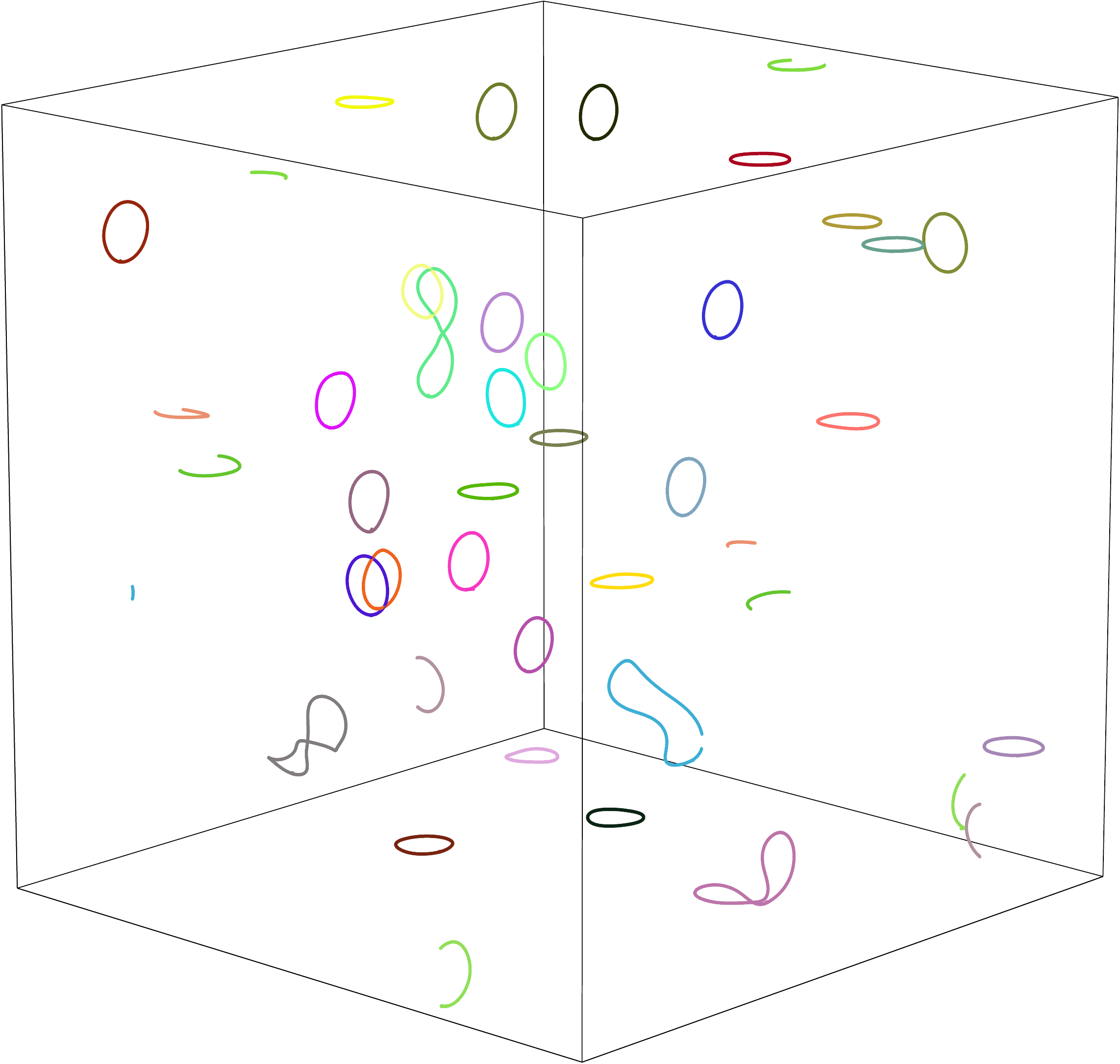}}
  \end{minipage}
  \caption{a) iso-surface plot corresponding to $|\psi|^2=0.1  \, |\Psi_\infty|^2  $ for a wave-function containing several rings, blue colours renders the sound waves; b) Plot of the vortex rings obtained by tracking zeros of the wave function $\psi$.
  \label{fig:Rings}}
\end{figure}

Having demonstrated the accuracy of the method, we now aim to test the robustness of the algorithm on more complex scenarios.
More specifically, we will show that the method is able to track complex vortex configurations characterized by several vortex loops.
As an example, we consider 20 vortex rings all having the same radii of $R=8\xi$ but randomly placed and oriented in the computational domain. To prepare the initial condition, we multiply 20 wave-functions (randomly translated and rotated) of a single vortex ring discretized on a grid with $256^3$ points and with a resolution $\Delta_x=\Delta_y=\Delta_z=\xi$.
We then evolve this initial condition for a short time before extracting the vortex positions to obtain a more realistic vortex configuration where also sound waves are present. 
In figures\ \ref{fig:Rings}(a) and \ref{fig:Rings}(b) we plot the iso-surfacess corresponding to $|\psi|^2=0.1  \, |\Psi_\infty|^2 $ and the tracked vortex loops, respectively.
Visually inspecting the two sub-plots provides qualitative confirmation that the method correctly detects
all the vortex rings present in the system. We remark that the computational time needed to track this configuration is less then one hour on a standard desktop machine.\\

\begin{figure}
 \begin{minipage}[b]{0.48\textwidth}
    \vspace{0pt}
    \centering
    \subfigure[]{
      \includegraphics[scale=0.22]{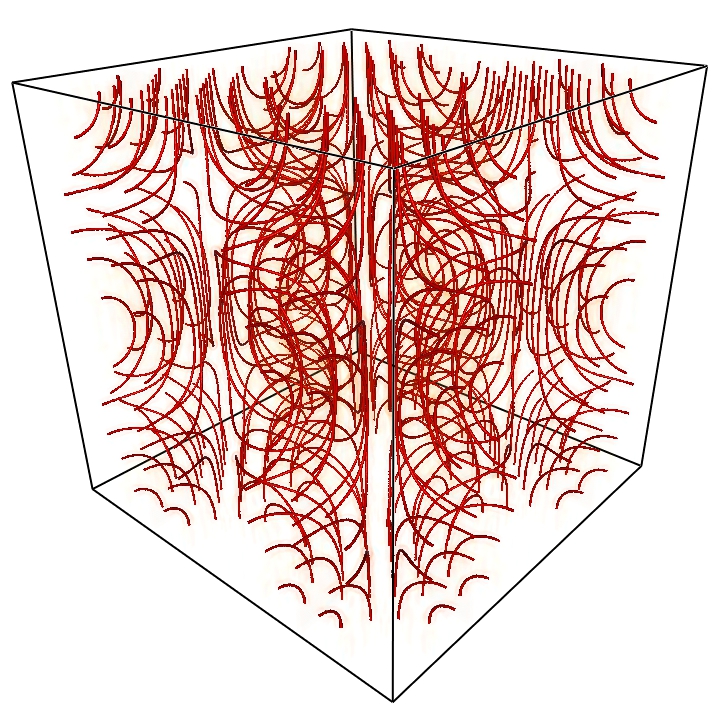}}
  \end{minipage}
 \begin{minipage}[b]{0.48\textwidth}
    \vspace{0pt} 
    \centering
    \subfigure[]{
          \includegraphics[scale=0.211]{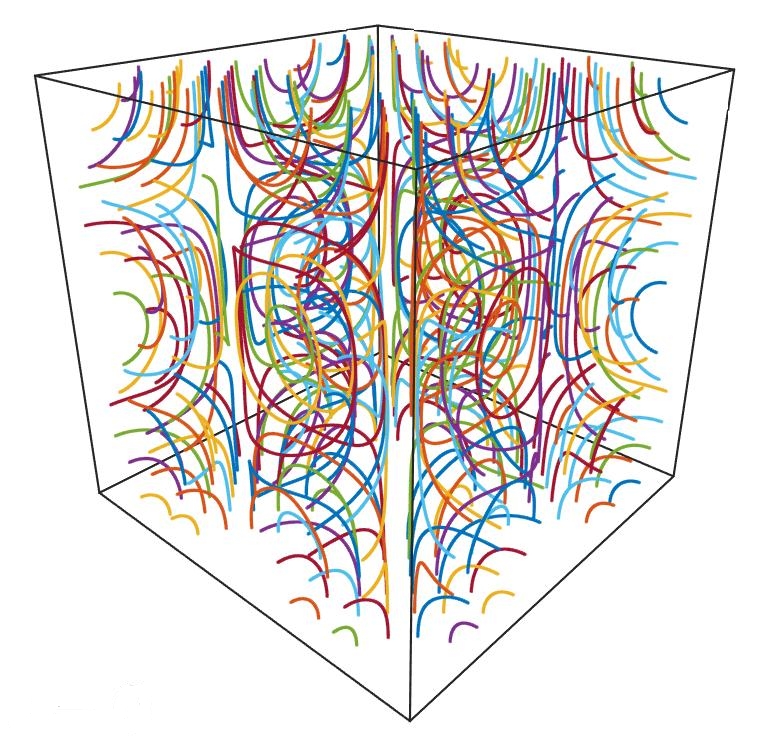}}
  \end{minipage}
  \begin{minipage}[b]{0.48\textwidth}
    \vspace{0pt}
    \centering
    \subfigure[]{
      \includegraphics[scale=0.22]{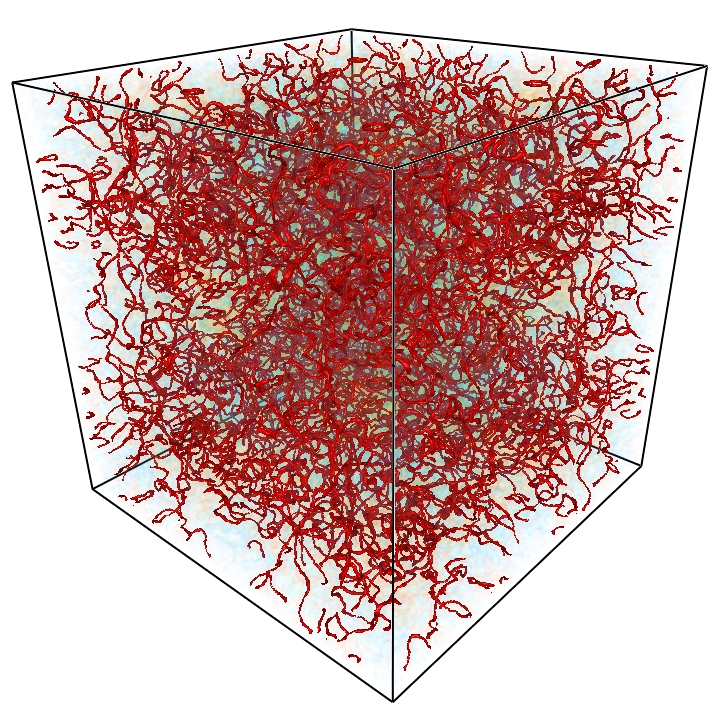}}
  \end{minipage}
 \begin{minipage}[b]{0.48\textwidth}
    \vspace{0pt} 
    \centering
    \subfigure[]{
          \includegraphics[scale=0.211]{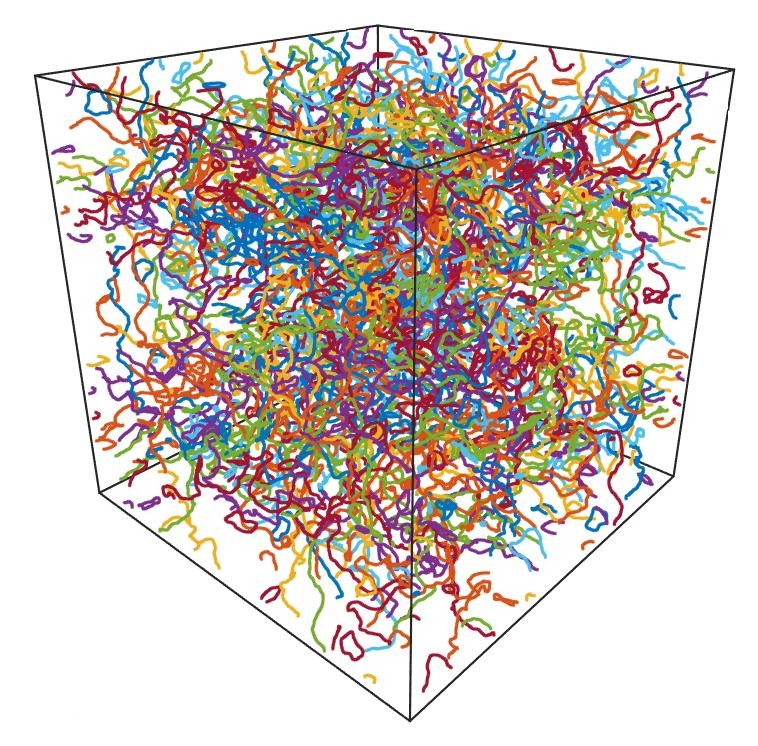}}
  \end{minipage}
  \caption{(Color online) Iso-surface plot corresponding to $|\psi|^2=0.1  \, |\Psi_\infty|^2 $ for the TG flow (a) at the initial condition and (c) after evolving the system for a time $t=10 \,t_{\mathrm{ed}}$, where $t_{\mathrm{ed}}$ corresponds to the largest eddy turnover time of the TG flow; tracked vortex filaments corresponding to (b) $t=0 \, t_{\mathrm{ed}}$ and (d) $t=10 \,t_{\mathrm{ed}}$. 
  }
  \label{fig:Tangle}
\end{figure}

For our final case study, we test the tracking method on a GP simulation containing a dense tangle of quantized vortices.
Such a configuration corresponds to a particularly important scenario in the study of quantum fluids since it represents an example of isotropic and homogeneous quantum turbulence. Due to the high density of vortices and the broad range of length scales involved in the dynamics, a vortex tangle also represents one of the most challenging vortex configurations to track.
Following the work of Nore {\em et al.}\cite{nore1997}, we created a tangle of vortices by evolving an initial configuration characterised by a so-called Taylor-Green (TG) flow. 
The iso-surfacess of the low-density field and the tracked vortex filaments corresponding to the initial condition are shown respectively in figures\ \ref{fig:Tangle}(a) and \ref{fig:Tangle}(b). 
We remark that the tracking has been carried out by setting $\zeta=\xi$ in equation\ (\ref{eq:rule}) and required a computational time of less than 1 hour on 64 cores running in parallel using MPI.
Upon integrating this initial condition forward in time with the GP equation, the interaction between the large scale vortex rings drives the system toward a vortex configuration characterised by a dense tangle where the superfluid motion occurs on a range of 
time and length scales within the system. This produces a quantum turbulent state as illustrated in the low density iso-surface plots shown in figure\ \ref{fig:Tangle}(c) that were obtained after a time $t \simeq 10 \, t_{\mathrm{ed}}$, where $t_{\mathrm{ed}}$ corresponds to the largest eddy turnover time of the TG flow.  In contrast to the initial condition, the tracking of such a complicated tangle requires us to set $\zeta=0.25\xi$ in order to explore all the small scales associated with the vortex dynamics. 
figure \ref{fig:Tangle}(d) shows the tracked vortex filaments; the tracking process took a computational time of approximately 6 hours on a cluster with 64 cores working in parallel using MPI.
Since we have no a-priori knowledge of the number of vortex rings composing the tangle displayed at $t=10 \, t_{\mathrm{ed}}$, no direct validation of the tracking is possible.

\begin{figure} 
  \begin{minipage}[b]{0.48\textwidth}
    \vspace{0pt}
    \centering
    \subfigure[]{
      \includegraphics[scale=0.37]{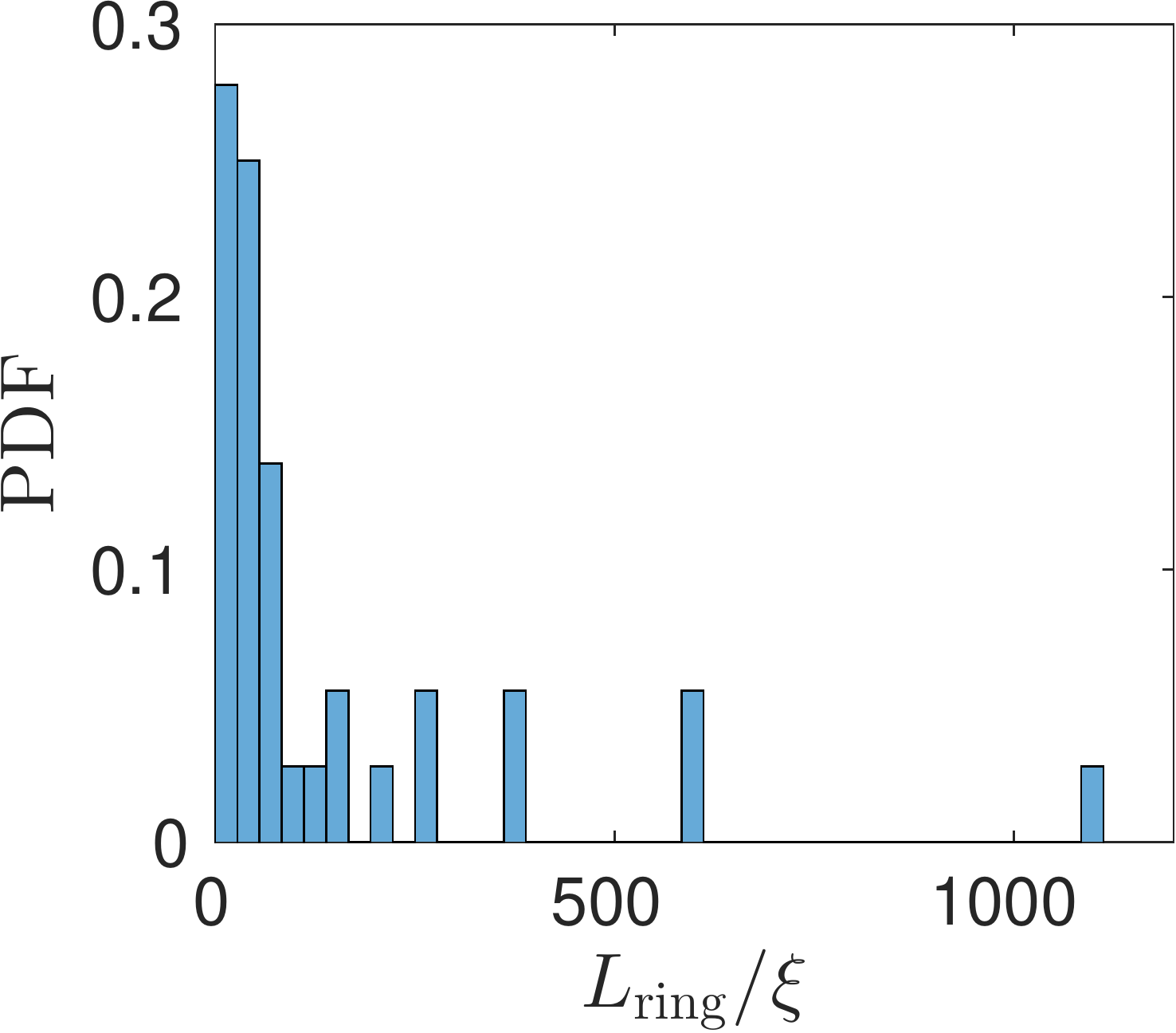}}
  \end{minipage}
   \begin{minipage}[b]{0.48\textwidth}
    \vspace{0pt} 
    \centering
    \subfigure[]{
          \includegraphics[scale=0.37]{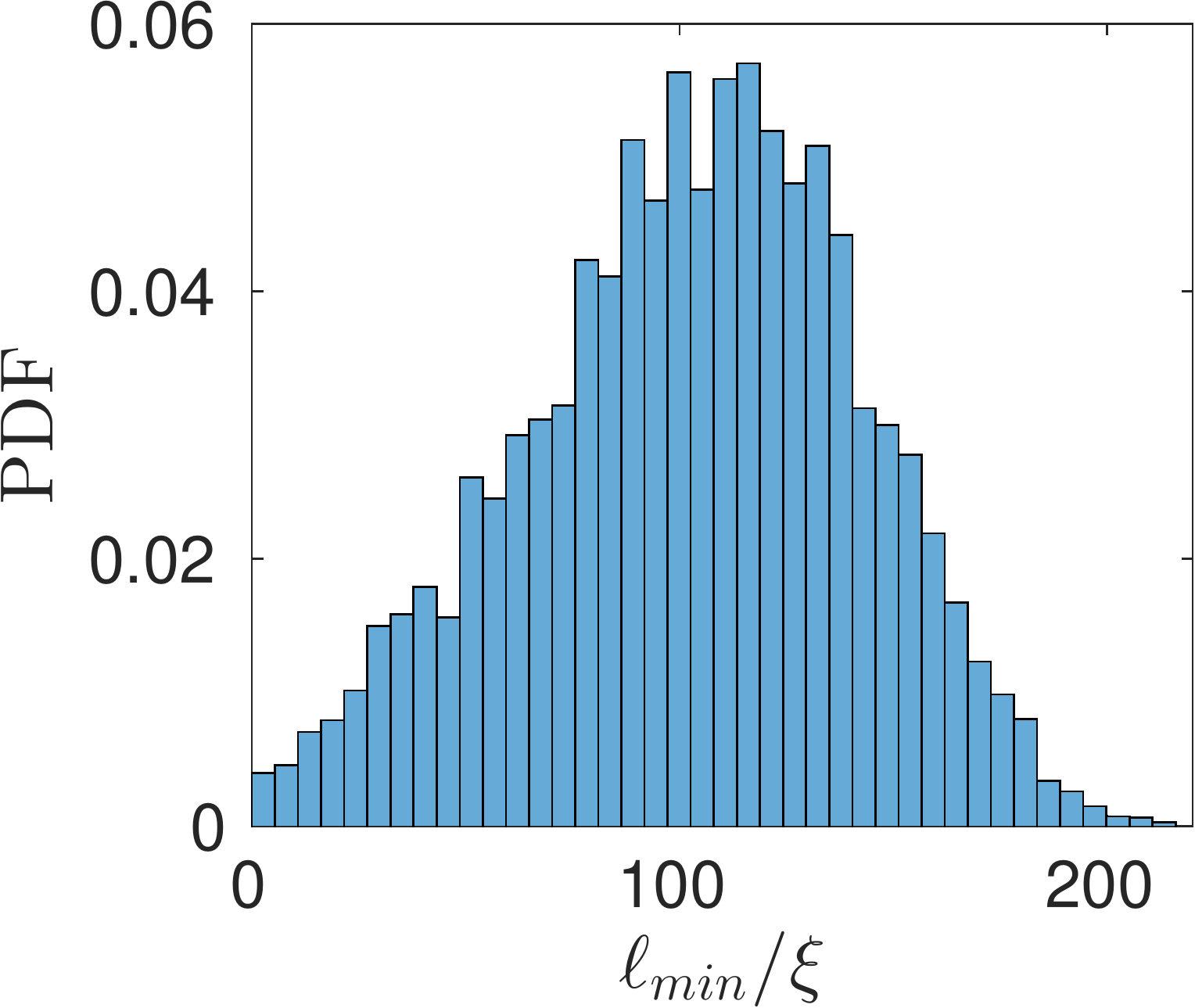}}
  \end{minipage}
  
   \hspace{0cm}
 \begin{minipage}[b]{0.48\textwidth}
    \vspace{0cm} 
    \centering
    \subfigure[]{
          \includegraphics[scale=0.36]{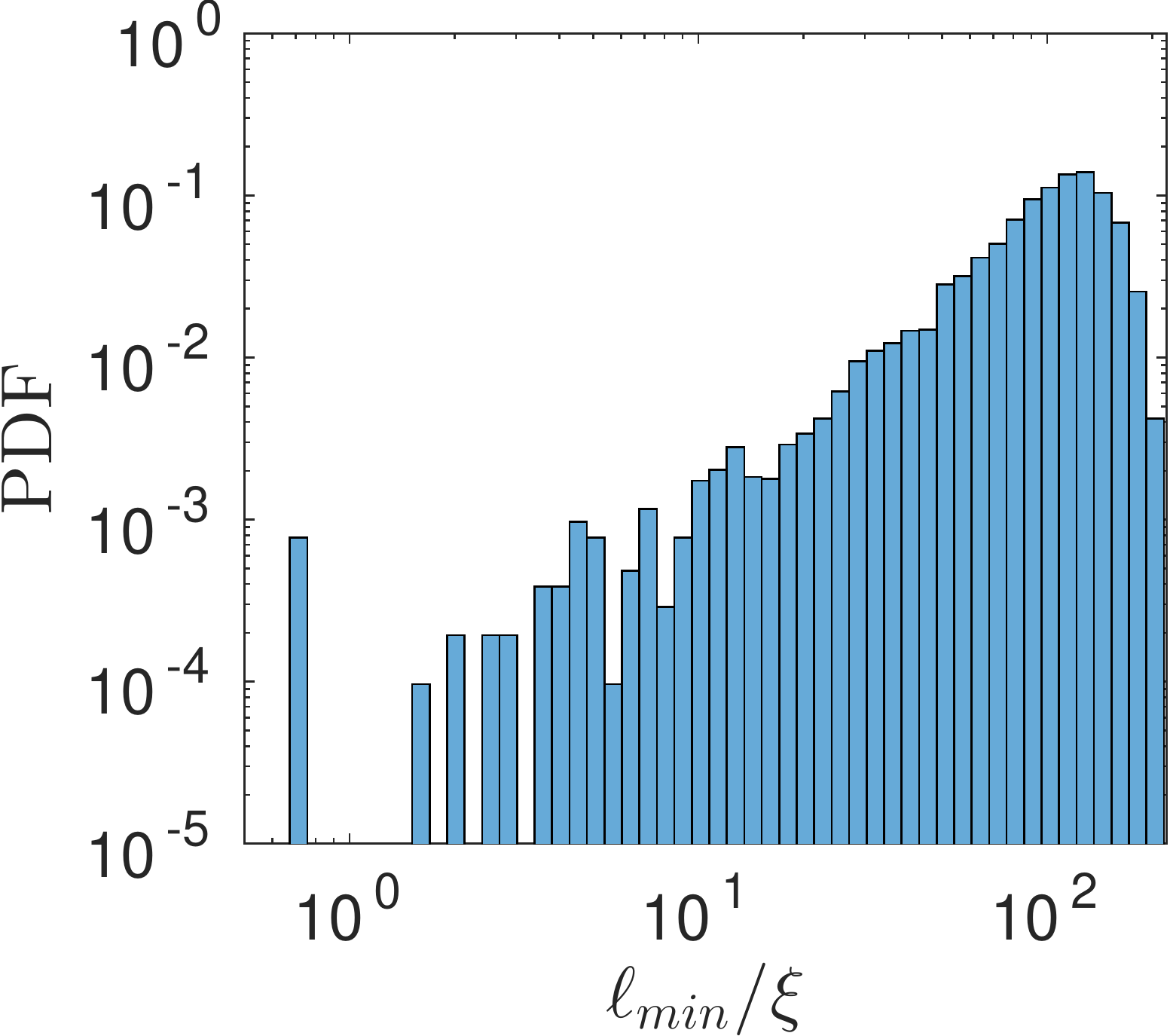}}
  \end{minipage}
  \hspace{1cm}
  \begin{minipage}[b]{0.48\textwidth}
    \subfigure[]{
      \includegraphics[scale=0.32]{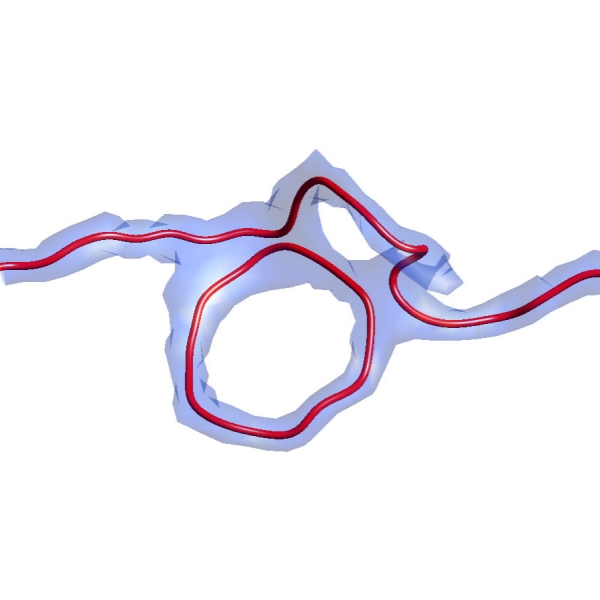}}
  \end{minipage}
  \caption{(Color online) (a) PDF plot of the sizes of the vortex rings; PDF plot of the minimal inter-vortex distance between each pair of vortex rings: (b) in linear scale and (c) in log-log scale; (d) Iso-surface plot (light blue) at $|\psi|^2=0.1  \, |\Psi_\infty|^2 $ for two vortices extracted from the tangle undergoing a reconnection. The tracked vortex positions are rendered with a red tube.\label{fig:PDF}}
\end{figure}

The tracking process reveals that the tangle is composed of $576$ rings with arc-lengths that vary from $L_{\mathrm{ring}}=8.4\xi$ to $L_{\mathrm{ring}}=1108\xi$. We point out that although the method should be capable of identifying even smaller vortex rings than those found, smaller rings tend to travel much faster and can shrink due to their interaction with sound waves that act as a thermal bath \cite{Krstulovic2011b}. Consequently, this can cause them to become rarefaction pulses according to the theory of Jones and Roberts\cite{Jones1982} which are not topological defects. This can explain the lack of smaller vortex rings within our simulations. A more complete analysis of the rings making up the vortex tangle is given by the probability distribution function (PDF) of the ring sizes displayed in figure\ref{fig:PDF}(a).
We see that most of the rings have relatively small size.
Such insight into the statistical properties of small vortices can help in estimating the amount of energy transferred into sound waves 
that compose the bath of thermal excitations through the process of vortex shrinking and annihilation. 
It is also true that the same PDF exhibits quite large fluctuations which indicates the presence of relatively large vortex rings.
Such large vortices are good candidates for testing predictions of KW spectra\cite{springerlink:10.1134,PhysRevLett.92.035301,PhysRevLett.91.135301, PhysRevB.85.104516} because they span a broad range of scales. 

Our tracking algorithm also allows us to explore the spatial distribution of the vortex tangle.
For instance, we can evaluate the minimal inter-vortex distance between the $ i $'th and $ j $'th vortex rings
\begin{equation}
\ell_{min}^{(i, j)} = \min \left( |\vv{x}^{(i)}-\vv{x}^{(j)}| \right) \, ,
\end{equation}
defined by choosing the minimum value of the distance between their vortex points.
The PDF of the minimal inter-vortex distances obtained by considering all the combinations of vortex ring pairs is displayed in figure \ref{fig:PDF}(b). The mean value of this distribution is given by $ \langle \ell_{min} \rangle = 102.7 \xi $.
Another important quantity to quantify in superfluid turbulence is the mean inter-vortex distance which is estimated to be $ \ell_{est} \sim 1/\sqrt{\mathcal{L}}$ where $\mathcal{L}$ is the total vortex length per unit volume\cite{donnelly1991}.
Using our tracking algorithm, we are able to precisely determine the total vortex line length which we find to be $ 80765.6 \xi$, providing an estimated mean inter-vortex distance of $ \ell_{est} = 14.4\xi$.
We can define the average inter-vortex distance between the $ i $'th and $ j $'th vortex rings as
\begin{equation}
\ell_{mean}^{(i, j)} = \frac{1}{\mathcal{N}_i \, \mathcal{N}_j} \sum_{k, l} |\vv{x}^{(i)}_k-\vv{x}^{(j)}_l| \, ,
\label{eq:meanEll}
\end{equation}
where $\mathcal{N}_i$ and $\mathcal{N}_j$ are the number of points contained in the $i$'th and $j$'th vortex rings, respectively. The tracked mean inter-vortex distance can then be evaluated as
\begin{equation}
\langle \ell \rangle_{track} = \frac{1}{n_v} \sum_{i=1}^{n_v} \min_{j\ne i} \left( \ell_{mean}^{(i, j)} \right) \, ,
\end{equation}
where $ n_v $ is the total number of vortices in the system.
In this particular case we obtain $ \langle \ell \rangle_{track} = 25.7\xi $ which is of the same order of $ \ell_{est} $.


Finally, in figure \ref{fig:PDF}(c) we plot the minimal inter-vortex distance in logarithmic scales in order to uncover the trend of the distribution at distances close to zero.
Remarkably, we notice that the algorithm is able to track vortex filaments whose minimal inter-vortex distance is below the computational grid ($ \Delta x= \Delta y= \Delta z= \xi $).
In figure \ref{fig:PDF}(d), we zoom in on a part of the tangle to reveal two vortices with a minimal inter-vortex separation of $ 0.6\xi $.
This figure further illustrates the accuracy of the algorithm and demonstrates the possibilities it provides in studying processes occurring on very small scales including vortex reconnection events that have generated significant interest in recent years\cite{zuccher2012,Rorai2014}.

\section{Conclusion and future perspectives}
In this work we have presented a numerical method to detect vortices of the complex wave-function $\psi$ describing a quantum fluid governed by the Gross--Pitaevskii model. 
The tracking is based on a Newton-Raphson root-finding algorithm to detect zeros of the complex field and employs the pseudo-vorticity field to track each vortex filament. 
We assumed the complex field is periodic that allowed us to make extensive use of spectral techniques to compute the field and its derivatives at any point in the computational domain. This assumption takes advantage of the spectral representation of the discretised field that is commonly used to integrate the GP equation resulting in an algorithm that is spectrally accurate.
Moreover, the method is formulated in a general way making no assumptions about the geometry or topology of the vortex filaments. It is, therefore, applicable to any vortex configuration that might arise in the dynamics of superfluid vortices.

We have presented several case studies to assess the accuracy and robustness our vortex tracking algorithm which are relevant to our understanding of the role of quantized vortices in superfluids. 
We demonstrated the accuracy of the method by showing that it can be used to accurately calculate the curvature and torsion, two quantities that are important in quantifying properties of superfluid vortices. 
We remark, that because of the spectral representation that we adopt for the complex wave-function, our method allows us to compute any high order directional derivatives along the vortex line by using the tangent vector along the filament (that can be computed from the pseudo-vorticity field) rather than using a spline interpolation scheme.
Moreover, we showed that it is capable of extracting information over a broad range of scales including small scale KW oscillations and to distinguish between filaments with an inter-vortex separation that is smaller than the computational grid size. This opens up the possibility to accurately study aspects of vortex reconnections in the GP model.


Precise knowledge of the location of vortex filaments is also useful in quantifying the topological complexity of a tangle since it allows us to evaluate important measures including, the linking, writhe, twist and helicity of the vortices. Complete knowledge of the positions of vortex filaments also allows us to extract statistical information regarding their properties. This can include the distribution of the sizes of vortex loops and the inter-vortex separations which can provide physical insights into the study of the dynamics and decay of the turbulent tangle. For instance, using this method, it is possible to study the evolution of the total vortex line length and to compute the velocity spectra associated with the vortex flow, both quantities that can be measured in quantum turbulence experiments.
It is also possible to obtain information about the emergence of spatio-temporal ordering, namely if vortices forming the tangle can organise themselves in oriented bundles or are composed of lines randomly oriented in space with respect to one another.

The method we have presented provides a broad scope for exploring many aspects of superfluid turbulence that have been hindered by the inability to obtain direct access to the information associated with the vortex filaments. It also provides a means to bridge the gap between vortex filament models and the GP model allowing us to test to what extent the models agree with one another. As a final remark, we note that by suitably extending the method presented in this work for the GP equation, it should be possible to track 
topological defects in other systems including nematic liquid crystals, superconductors, skyrmions, cosmic strings and optical vortices. Although some of these systems can exhibit a more complex and diverse range of topological defects, the principle approach should be generalizable to such systems. This would contribute to our understanding of the role of topological defects across a range of condensed matter systems.

\ack
GK, DP and AV were supported by the cost-share Royal Society International Exchanges Scheme (reference IE150527) in conjunction with CNRS. HS acknowledges support for a Research Fellowship from the Leverhulme Trust under Grant R201540.
Computations were carried out on the M\'esocentre SIGAMM hosted at the Observatoire de la C\^ote d'Azur and on the High Performance Computing Cluster supported by the Research and Specialist Computing Support service at the University of East Anglia.

\appendix

\section{Wave-function of a torus vortex knot\label{App:Knot}}

For completeness, we present here the wave-function used for producing the knot displayed in figure \ref{fig:Knot}. It is based on the formula introduced in Proment {\em et al.} \cite{PhysRevE.85.036306} and given by
\begin{eqnarray}
\fl \psi_{p, q}(x, y, z) & = & \prod_{i=1}^{p} \Psi_{2D}\left\{ s(x, y) - R_0 - R_1 \cos \left[\alpha(x, y) + i \frac{2\pi \,q}{p}\right], z - R_1 \sin \left[\alpha(x, y) + i \frac{2\pi \,q}{p}\right] \right\}
\nonumber \\ 
\fl  & \times & 
\prod_{i=1}^{p} \Psi_{2D}^\ast \left\{ s(x, y) + R_0 + R_1 \cos \left[\alpha(x, y) + i \frac{2\pi \,q}{p}\right], z - R_1 \sin \left[\alpha(x, y) + i \frac{2\pi \,q}{p}\right] \right\} \nonumber \\
\fl &&
\label{eq:Tpq}
\end{eqnarray}
where
\begin{equation}
s(x, y)=\mbox{sgn}(x) \, \sqrt{x^2+y^2}, \quad \mbox{and} \quad \alpha(x, y)=\frac{q \, \mbox{atan2}(x, y)}{p} \, .
\end{equation}
Here, $ \Psi_{2D}(s-s_0, z-z_0) $ describes the two-dimensional wave-function of a single vortex centred at $ (s_0, z_0) $\cite{Berloff2004}, $ \mbox{sgn}(\cdot) $ is the sign function, and $ \mbox{atan2}(\cdot, \cdot) $ is the four-quadrant arctangent. Starred quantities, $ (\cdot)^\ast $, denote the complex conjugate. In order to account for the periodic boundary conditions assumed in this work, we modify equation\ (\ref{eq:Tpq}) by setting
\begin{eqnarray}
\psi_{p,q }^{(per)}(x,y,z)&=&\psi_{p, q}(x_p,y_p,z_p), \label{eq:psi_per_app}
\end{eqnarray}
where $x_p,y_p$ and $z_p$ is a periodic approximation of the identity at $(L/2,L/2,L/4)$. Namely
\begin{equation}
x_p= -\frac{L}{\pi} \cos{\frac{\pi}{L}x},\hspace{.5cm}y_p= -\frac{L}{\pi} \cos{\frac{\pi}{L}y},\hspace{.5cm} z_p=-\frac{L}{2 \pi}\cos{\frac{2\pi}{L}z}. \label{Eq:ApproxIdententyPeriodic}
\end{equation}
Torsion and curvature displayed in figures\ \ref{fig:Knot}(c) and \ref{fig:Knot}(d) are then computed by accounting for 
corrections due to the assumed periodicity of the field according to equation\ (\ref{Eq:ApproxIdententyPeriodic}).
The knot is thus defined by the line 
\[
(x_p^{-1}(s_x(\sigma)),y_p^{-1}(s_y(\sigma)),z_p^{-1}(s_z(\sigma)),
\] 
obtained from equation\ (\ref{Eq:Knot}).

\section{Definition of Kelvin wave spectra\label{App:KW}}

Kelvin waves are helical oscillations propagating along vortex lines. In this work, we have studied KWs on straight lines and rings. One of the simplest two-point quantities of interest is the KW spectrum, that gives the amplitude of KWs at a given scale. The definition of the spectra slightly differs for lines and rings. 

A straight line with small amplitude KWs, can be easily parametrised as
\begin{equation}
{\bf S}_{\rm KW}(\sigma)=(X(\sigma),Y(\sigma),\sigma).
\end{equation}
Assuming that the line is periodic in the $z$ direction, the KW spectrum is  defined as
\begin{equation}
n_k=|\widehat{{\bf S}_{\rm KW}}(k)|^2+|\widehat{{\bf S}_{\rm KW}}(-k)|^2,\label{Eq:kwSp}
\end{equation}
where $\widehat{{\bf S}_{\rm KW}}(k)$ is the Fourier transform of ${\bf S}_{\rm KW}(\sigma)=(X(\sigma),Y(\sigma))$.
Note that lines obtained from the tracking are not directly parametrised in terms of $\sigma$. In order to numerically compute the Fourier transform, the line is remeshed on a regular partition of the interval $[0,L_z]$ (where $L_z$ is the size of the periodic domain along the $z$-coordinate direction) obtained by a high-order interpolation.  In Krstulovic \cite{Krstulovic2012}, KWs on a straight line were accurately tracked by a NR method using planes perpendicular to the line. The algorithm used in \cite{Krstulovic2012} provided a parametrization on a regular mesh that directly allows for the evaluation of the Fourier transform. We have checked that the interpolation does not affect the values of the spectrum at the scales of interest.

\begin{figure}
\center
\includegraphics[width=0.75\textwidth]{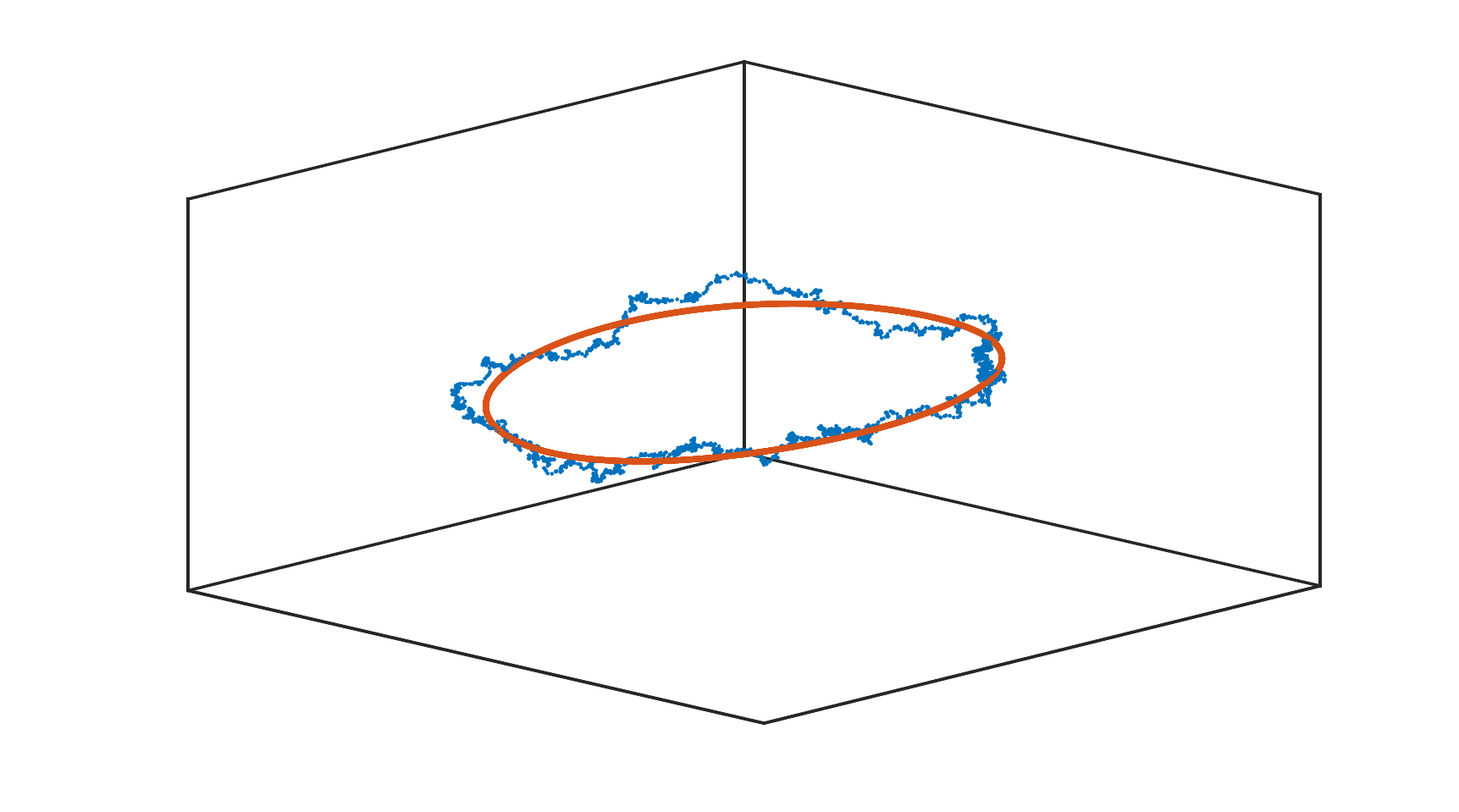}
\caption{(Color online) Smooth (red line) and perturbed vortex ring with imposed KW spectrum (blue line).}
\label{Fig:KWringTest}
\end{figure}

In order to obtain the KW spectrum of waves on a ring we proceed as follows. Starting from a ring with Kelvin waves ${\bf S}_{\rm KW}(s)$ expressed in its natural parametrisation, we obtain a long-scale averaged ring ${\bf S}_{\rm smooth}(s)$ by convolving the line with a Gaussian kernel of width $\alpha L$, where $\alpha\in(0,1)$ and $L$ is the total length of the line. Once this smooth ring is computed, we define the KWs as 
\begin{equation}
{\bf S}_{\rm KW}(s)={\bf S}(s)-{\bf S}_{\rm smooth}(s).
\end{equation}
Figure \ref{Fig:KWringTest} displays a test case with KWs (shown in blue) and the corresponding smooth ring (shown in red). 
By construction, ${\bf S}_{\rm KW}(s)$ is a periodic set of 3 signals (one for each spatial dimension). The KW spectrum is then simply defined as in (\ref{Eq:kwSp}) using the Fourier transform of ${\bf S}_{\rm KW}(s)$.

Note that the tracking algorithm provides a parametrisation $\sigma$ that is close to the natural one but not necessarily equal as after each step $\zeta$, the location of the vortex is re-evaluated in the new plane so the distance between the previous point and the new one can be slightly different from $\zeta$. As in the case of KWs on a straight vortex, the line obtained from the algorithm is reparametrized in a regular partition of $[0,L]$ by using a high order interpolation. In this work we used a value of $\alpha=0.1$. Varying this coefficient only slightly modifies the large-scale values of the spectrum, but values at scales where scaling is observed remain unchanged.

\bibliographystyle{ieeetr}

\bibliography{Ref_DP}

\begin{thebibliography}{10}

\bibitem{donnelly1991}
R.~Donnelly, {\em {Quantized vortices in Helium II}}, vol.~3.
\newblock Cambridge Univ Pr, 1991.

\bibitem{pitaevskii2003bose}
L.~Pitaevskii and S.~Stringari, {\em {Bose-Einstein Condensation}}, vol.~116.
\newblock Oxford University Press, USA, 2003.

\bibitem{Schwarz1985}
K.~W. Schwarz, ``Three-dimensional vortex dynamics in superfluid
  $^{4}\mathrm{He}$: Line-line and line-boundary interactions,'' {\em Phys.
  Rev. B}, vol.~31, pp.~5782--5804, May 1985.

\bibitem{Saffman1992}
P.~Saffman, {\em Vortex Dynamics}.
\newblock Cambridge University Press, 1993.
\newblock Cambridge Books Online.

\bibitem{Roberts2001}
P.~Roberts and N.~Berloff, ``{The Nonlinear Schr\"odinger Equation as a Model
  of Superfluidity},'' in {\em Quantized Vortex Dynamics and Superfluid
  Turbulence} (C.~Barenghi, R.~Donnelly, and W.~Vinen, eds.), vol.~571 of {\em
  Lecture Notes in Physics}, pp.~235--257, Springer Berlin Heidelberg, 2001.

\bibitem{koplik1993vrs}
J.~Koplik and H.~Levine, ``{Vortex reconnection in superfluid helium},'' {\em
  Physical Review Letters}, vol.~71, no.~9, pp.~1375--1378, 1993.

\bibitem{nazarenko2003}
S.~Nazarenko and R.~West, ``{Analytical Solution for Nonlinear Schr\"odinger
  Vortex Reconnection},'' {\em Journal of Low Temperature Physics}, vol.~132,
  no.~1, pp.~1--10.

\bibitem{zuccher2012}
S.~Zuccher, M.~Caliari, A.~W. Baggaley, and C.~F. Barenghi, ``Quantum vortex
  reconnections,'' {\em Physics of Fluids}, vol.~24, no.~12, 2012.

\bibitem{Krstulovic2012}
G.~Krstulovic, ``Kelvin-wave cascade and dissipation in low-temperature
  superfluid vortices,'' {\em Phys. Rev. E}, vol.~86, p.~055301, Nov 2012.

\bibitem{2013arXiv1308.0852P}
D.~{Proment}, C.~F. {Barenghi}, and M.~{Onorato}, ``{Interaction and decay of
  Kelvin waves in the Gross-Pitaevskii model},'' {\em ArXiv e-prints}, Aug.
  2013.

\bibitem{diLeoniKW}
P.~Clark~di Leoni, P.~D. Mininni, and M.~E. Brachet, ``{Spatiotemporal
  detection of Kelvin waves in quantum turbulence simulations},'' {\em Phys.
  Rev. A}, vol.~92, p.~063632, Dec 2015.

\bibitem{Vinen2001}
W.~F. Vinen, ``{Decay of superfluid turbulence at a very low temperature: The
  radiation of sound from a Kelvin wave on a quantized vortex},'' {\em Phys.
  Rev. B}, vol.~64, p.~134520, Sep 2001.

\bibitem{Scheeler2014}
M.~W. Scheeler, D.~Kleckner, D.~Proment, G.~L. Kindlmann, and W.~T.~M. Irvine,
  ``Helicity conservation by flow across scales in reconnecting vortex links
  and knots,'' {\em Proceedings of the National Academy of Sciences}, vol.~111,
  no.~43, pp.~15350--15355, 2014.

\bibitem{PhysRevE.92.061001}
S.~Zuccher and R.~L. Ricca, ``Helicity conservation under quantum reconnection
  of vortex rings,'' {\em Phys. Rev. E}, vol.~92, p.~061001, Dec 2015.

\bibitem{Brachet2016}
P.~{Clark di Leoni}, P.~D. {Mininni}, and M.~E. {Brachet}, ``{Helicity,
  Topology and Kelvin Waves in reconnecting quantum knots},'' {\em ArXiv
  e-prints}, Feb. 2016.

\bibitem{nore1997}
C.~Nore, M.~Abid, and M.~Brachet, ``{Decaying Kolmogorov turbulence in a model
  of superflow},'' {\em Physics of Fluids (1994-present)}, vol.~9, no.~9,
  pp.~2644--2669, 1997.

\bibitem{berloff2002ssn}
N.~Berloff and B.~Svistunov, ``{Scenario of strongly nonequilibrated
  Bose-Einstein condensation},'' {\em Physical Review A}, vol.~66, no.~1,
  p.~13603, 2002.

\bibitem{yepez:084501}
J.~Yepez, G.~Vahala, L.~Vahala, and M.~Soe, ``Superfluid turbulence from
  quantum kelvin wave to classical kolmogorov cascades,'' {\em Physical Review
  Letters}, vol.~103, no.~8, p.~084501, 2009.

\bibitem{Taylor2014}
A.~J. Taylor and M.~R. Dennis, ``Geometry and scaling of tangled vortex lines
  in three-dimensional random wave fields,'' {\em Journal of Physics A:
  Mathematical and Theoretical}, vol.~47, no.~46, p.~465101, 2014.

\bibitem{PhysRevA.69.053601}
N.~G. Berloff, ``Interactions of vortices with rarefaction solitary waves in a
  bose-einstein condensate and their role in the decay of superfluid
  turbulence,'' {\em Phys. Rev. A}, vol.~69, p.~053601, May 2004.

\bibitem{Rorai2014}
C.~{Rorai}, J.~{Skipper}, R.~M. {Kerr}, and K.~R. {Sreenivasan}, ``{Approach
  and separation of quantum vortices with balanced cores},'' {\em ArXiv
  e-prints}, Oct. 2014.

\bibitem{NumRecipes}
W.~H. Press, S.~A. Teukolsky, W.~T. Vetterling, and B.~P. Flannery, {\em
  Numerical Recipes 3rd Edition: The Art of Scientific Computing}.
\newblock New York, NY, USA: Cambridge University Press, 3~ed., 2007.

\bibitem{Krstulovic2DVortices}
G.~Krstulovic, M.~Brachet, and E.~Tirapegui, ``{Radiation and vortex dynamics
  in the nonlinear Schr\"odinger equation},'' {\em Phys. Rev. E}, vol.~78,
  p.~026601, Aug 2008.

\bibitem{pismen1999vortices}
L.~M. Pismen, {\em Vortices in nonlinear fields: From liquid crystals to
  superfluids, from non-equilibrium patterns to cosmic strings}, vol.~100.
\newblock Oxford University Press, 1999.

\bibitem{Berloff2004}
N.~Berloff, ``{Pad{\'e} approximations of solitary wave solutions of the
  Gross-Pitaevskii equation},'' {\em Journal of Physics A: Mathematical and
  General}, vol.~37, no.~5, p.~1617, 2004.

\bibitem{Salman2013}
H.~Salman, ``{Breathers on Quantized Superfluid Vortices},'' {\em Phys. Rev.
  Lett.}, vol.~111, p.~165301, 2013.

\bibitem{KondaurovaVFMStat}
L.~Kondaurova, V.~L'vov, A.~Pomyalov, and I.~Procaccia, ``{Structure of a
  quantum vortex tangle in ${}^{4}$He counterflow turbulence},'' {\em Phys.
  Rev. B}, vol.~89, p.~014502, Jan 2014.

\bibitem{maggioni:2010}
F.~Maggioni, S.~Alamri, C.~F. Barenghi, and R.~L. Ricca, ``Velocity, energy,
  and helicity of vortex knots and unknots,'' {\em Physical Review E}, vol.~82,
  p.~026309, 2010.

\bibitem{PhysRevE.85.036306}
D.~Proment, M.~Onorato, and C.~F. Barenghi, ``{Vortex knots in a Bose-Einstein
  condensate},'' {\em Phys. Rev. E}, vol.~85, p.~036306, Mar 2012.

\bibitem{1742-6596-544-1-012022}
D.~Proment, M.~Onorato, and C.~F. Barenghi, ``{Torus quantum vortex knots in
  the Gross-Pitaevskii model for Bose-Einstein condensates},'' {\em Journal of
  Physics: Conference Series}, vol.~544, no.~1, p.~012022, 2014.

\bibitem{Kleckner:2016yu}
D.~Kleckner, L.~H. Kauffman, and W.~T.~M. Irvine, ``How superfluid vortex knots
  untie,'' {\em Nat Phys}, vol.~advance online publication, pp.~--, 03 2016.

\bibitem{Hasimoto1972}
H.~Hasimoto, ``A soliton on a vortex filament,'' {\em Journal of Fluid
  Mechanics}, vol.~51, pp.~477--485, 2 1972.

\bibitem{Ricca1992}
R.~L. Ricca and H.~Moffatt, {\em Topological Aspects of the Dynamics of Fluids
  and Plasmas}, ch.~The helicity of a knotted vortex filament.
\newblock Springer, 1992.

\bibitem{Hietala2016}
N.~Hietala, R.~H\"anninen, and H.~Salman, ``Helicity and internal twist within
  the vortex filament model,'' {\em arXiv:1604.03350}, 2016.

\bibitem{Krstulovic2011b}
G.~Krstulovic and M.~Brachet, ``Energy cascade with small-scale thermalization,
  counterflow metastability, and anomalous velocity of vortex rings in
  fourier-truncated gross-pitaevskii equation,'' {\em Physical Review E},
  vol.~83, no.~6, p.~066311, 2011.

\bibitem{Jones1982}
C.~Jones and P.~Roberts, ``{Motions in a Bose condensate: IV. Axisymmetric
  solitary waves},'' {\em J.\ Phys.\ A:Math.\ Gen.}, vol.~15, pp.~2599--2619,
  1982.

\bibitem{springerlink:10.1134}
V.~L'vov and S.~Nazarenko, ``{Spectrum of Kelvin-wave turbulence in
  superfluids},'' {\em JETP Letters}, vol.~91, pp.~428--434, 2010.
\newblock 10.1134/S002136401008014X.

\bibitem{PhysRevLett.92.035301}
E.~Kozik and B.~Svistunov, ``Kelvin-wave cascade and decay of superfluid
  turbulence,'' {\em Phys. Rev. Lett.}, vol.~92, p.~035301, Jan 2004.

\bibitem{PhysRevLett.91.135301}
W.~F. Vinen, M.~Tsubota, and A.~Mitani, ``Kelvin-wave cascade on a vortex in
  superfluid $^{4}\mathrm{H}\mathrm{e}$ at a very low temperature,'' {\em Phys.
  Rev. Lett.}, vol.~91, p.~135301, Sep 2003.

\bibitem{PhysRevB.85.104516}
E.~B. Sonin, ``Symmetry of kelvin-wave dynamics and the kelvin-wave cascade in
  the $t=0$ superfluid turbulence,'' {\em Phys. Rev. B}, vol.~85, p.~104516,
  Mar 2012.

\end{thebibliography}

\end{document}